\documentclass[aps,pre,twocolumn,superscriptaddress,noshowpacs]{revtex4}
\usepackage{amsmath}
\usepackage{amssymb}
\usepackage{tikz}
\usepackage{graphicx}
\usepackage{dcolumn}
\usepackage{bm}
\usepackage{epsfig}
\usepackage{psfrag}
\usepackage{latexsym}

\newcommand{\reffig}[1]{\mbox{Fig.~\ref{#1}}}
\newcommand{\refeq}[1]{\mbox{Eq.~(\ref{#1})}}
\newcommand{\refsec}[1]{\mbox{Sec.~\ref{#1}}}

\newcommand{\be}{\begin{equation}}
\newcommand{\ee}{\end{equation}}
\newcommand{\ba}{\begin{eqnarray}}
\newcommand{\ea}{\end{eqnarray}}
\renewcommand{\Re}{\mathrm{Re}}

\newcommand{\T}{${\mathcal T}\,$}
\newcommand{\Ti}{${\mathcal T}$}

\usepackage{color}

\begin{document}
\draft
\title{\bf Missing level statistics in a dissipative microwave resonator with partially violated time-reversal invariance}
\author{Ma{\l}gorzata Bia{\l}ous}
\affiliation{%
Institute of Physics, Polish Academy of Sciences, Aleja Lotnik\'{o}w 32/46, 02-668 Warszawa, Poland
}
\author{Barbara Dietz}
\altaffiliation{
email:  Dietz@lzu.edu.cn}
\affiliation{%
Lanzhou Center for Theoretical Physics and the Gansu Provincial Key
Laboratory of Theoretical Physics, Lanzhou University,
Lanzhou, Gansu 730000, China
}
\author{Leszek Sirko}
\altaffiliation{
email:  sirko@ifpan.edu.pl}
\affiliation{%
Institute of Physics, Polish Academy of Sciences, Aleja Lotnik\'{o}w 32/46, 02-668 Warszawa, Poland
}
\date{\today}
\bigskip
\begin{abstract}
We report on the experimental investigation of the fluctuation properties in the resonance frequency spectra of a flat resonator simulating a dissipative quantum billiard subject to partial time-reversal invariance violation (TIV) which is induced by two magnetized ferrites. The cavity has the shape of a quarter bowtie billiard of which the corresponding classical dynamics is chaotic. Due to dissipation it is impossible to identify a complete list of resonance frequencies. Based on a random-matrix theory approach we derive analytical expressions for statistical measures of short- and long-range correlations in such incomplete spectra interpolating between the cases of preserved time-reversal invariance and complete TIV and demonstrate their applicability to the experimental spectra.
\end{abstract}

\bigskip
\maketitle
\smallskip
\section{Introduction}
The conjecture~\cite{Berry1979,Casati1980,Bohigas1984} that the fluctuation properties in the eigenvalue spectra of typical quantum systems with fully chaotic classical limit coincide with those of random matrices from the Gaussian ensembles~\cite{Mehta2004} has become the cornerstone of many theoretical, experimental and numerical studies in the field of quantum chaos. The spectral properties of generic quantum systems with fully chaotic classical dynamics and preserved time-reversal (\Ti) invariance coincide with those of random matrices from the Gaussian orthogonal ensemble (GOE). Some examples are quantum wells~\cite{Vina1998}, molecular spectra~\cite{Zimmermann1988}, atoms in a strong microwave field~\cite{Sirko1993,Sirko2002}, flat microwave resonators (billiards)~\cite{Stoeckmann1990,Graef1992,Sridhar1994,Hlushchuk2000,Hemmady2005,Hul2005,Dietz2015} and microwave networks~\cite{Hul2004,Hul2012,Bialous2017,Lawniczak2019}. In the presence of \Ti-invariance violation (TIV) the spectral properties are well described by those of random matrices from the Gaussian unitary ensemble (GUE). This was observed, for example, for atoms in a constant external magnetic fields~\cite{Sacha1999}, graphene quantum dots~\cite{Ponomarenko2008}, Rydberg excitons~\cite{Amann2016}, nuclear reactions~\cite{French1985,Mitchell2010}, microwave billiards~\cite{So1995,Stoffregen1995,Dietz2007a} and networks~\cite{Lawniczak2010,Bialous2016,Rehemanjiang2018,Lawniczak2019b,Lu2020,Yunko2020,Lawniczak2020}. A random-matrix theory (RMT) approach was also developed for quantum systems with partially violated \Ti-invariance~\cite{Pandey1991,Dietz1991,Lenz1992,Schierenberg2012}. Such systems were realized experimentally by inserting ferritic material into a microwave billiards and magnetizing it with an external magnetic field~\cite{So1995,Stoffregen1995,Dietz2007a,Dietz2009a}. The degree of time irreversibility can be adjusted by controlling the magnitude of the applied magnetic field and quantified by a parameter which interpolates between the cases of a \Ti-invariant system and a system with complete TIV~\cite{Pandey1991,Lenz1992,Dietz2019,Bialous2020,Yunko2020}. We would like to mention that dissipative microwave billiards subject to partial TIV were previously used in a different context in Refs.~\cite{Dietz2011,Dietz2012} to investigate the features of dissipative quantum systems with partial TIV in the vicinity of exceptional points, that is, those of isolated pairs of nearly-degenerate resonance frequencies. Furthermore, the fluctuation properties in the level sequence of a flat microwave resonators or quantum billiard are determined by their shape. Therefore, billiards are often used for the numerical, theoretical and experimental investigation of the features of quantum chaos.

Generally, the completeness of a spectrum is indispensable for the comparison of the fluctuation properties with those of random matrices from the Gaussian ensembles. However, complete sequences of several hundreds of levels are achievable only very rarely. One example are high-precision measurements of resonance spectra performed with flat, superconducting resonators at liquid-helium temperature~\cite{Dietz2014,Dietz2015,Dietz2016,Dietz2019} yielding sharp and well isolated resonances, and thus making the determination of the resonance frequencies from the positions of the resonances feasible. Due to absorption measurements at room temperature typically yield overlapping resononaces, which render the determination of complete sequences of resonances impossible. Thus, experimentally determined level sequences are typically incomplete and lead to deviations of the spectral properties from RMT predictions for complete spectra. To overcome these difficulties one has to cope with missing levels, which has been made feasible with the RMT approach for missing levels introduced in Ref.~\cite{Bohigas2004}. It was developed in the context of nuclear physics in Refs.~\cite{Agvaanluvsan2003,Agvaanluvsan2003a,Bohigas2004,Bohigas2006}  and applied to atomic, molecular and nuclear systems~\cite{Liou1972,Brody1981,Zimmermann1988,Enders2000,Enders2004,Molina2007,Frisch2014,Mur2015} and to microwave billiards and microwave networks~\cite{Bialous2016,Bialous2016a,Lawniczak2018} simulating quantum billiards and graphs with preserved \T invariance and complete TIV.

The objective of this article is to analyze the spectral fluctuations in the resonance frequency spectra of a real system, a microwave resonator with internal absorption which leads to the incompleteness of the spectra, in the presence of partial TIV. We present analytical results accounting for partial TIV in terms of a parameter $\xi$ \cite{Lenz1992} which interpolates between GOE and GUE and for the incompleteness of the level sequence which is characterized by the fraction $\Phi$ of identified levels. In~\refsec{Exp} we describe the experimental setup and then introduce in~\refsec{Missing} the RMT approach for missing levels and apply it to the experimental level sequences. In~\refsec{Concl} we summarize the results.

\section{Experimental setup\label{Exp}}
\begin{figure}[tb]
\begin{center}
\includegraphics[width=\linewidth]{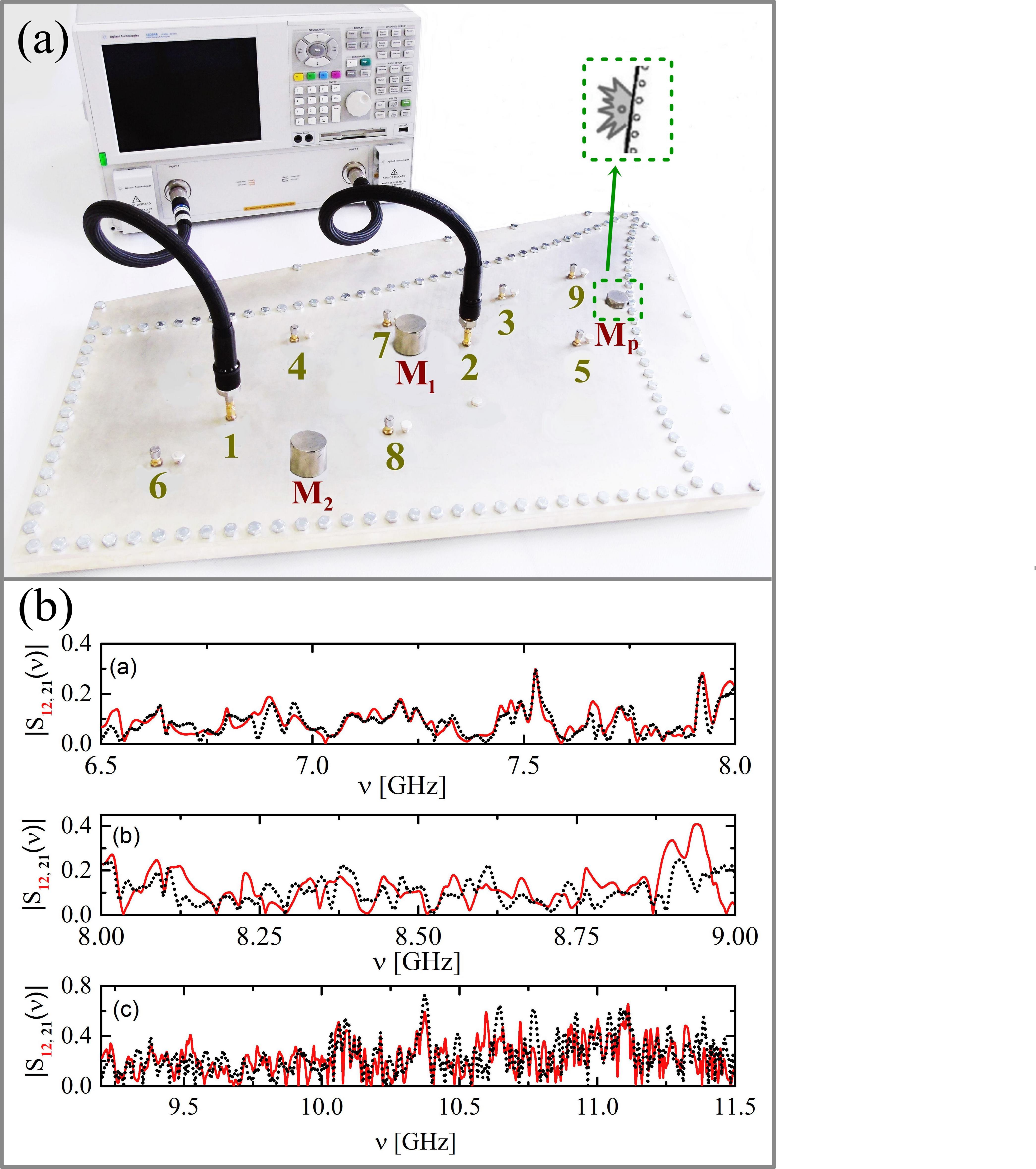}
	\caption{ (a) The experimental setup. The vector network analyzer Agilent E8364B is connected to the microwave antennas, which are attached to the resonator through the flexible microwave cables. In order to induce $\mathcal{T}$-invariance violation two pieces of ferrite are inserted into the cavity and magnetized by four extarnal magnets placed at the positions of the ferrites below and above the resonator. The latter are marked by M$_1$ and M$_2$. An additional magnet M$_P$ is used to move a metallic perturber inside the cavity alongside its walls to create different realizations of the cavity. (b) Transmission spectra from antenna 1 to antenna 2 (black) and vice versa (green) in three frequency regions.
}\label{Fig1}
\end{center}
\end{figure}

We used the same microwave cavity (see~\reffig{Fig1}) as in our previous work on the enhancement factor as function of openness and size of TIV~\cite{Bialous2019,Bialous2020}. It has the shape of a quarter bowtie billiard of which the classical dynamics is fully chaotic. The cavity consists of two plates of polished aluminum type EN 5754. A basin of area $\mathcal{A}=1828.5 \pm 5.0$ cm$^2$, perimeter $\mathcal{L}=202.3 \pm 2.0$ cm and depth $h=1.2$~cm, which forms the resonator body, was milled out of the bottom plate. The inner surface of the cavity is covered with a 20 $\mu$m layer of silver to reduce internal absorption. Below the cut-off frequency of $\nu_{max}=c/2h \simeq 12.49$ GHz, with $c$ denoting the speed of light in vacuum, only the transverse magnetic modes are excited inside the cavity so that the Helmholtz equation describing the electromagnetic field in the microwave cavity and the two-dimensional Schr\"odinger equation for the quantum billiard of corresponding shape are mathematically equivalent. The top plate of the cavity has nine identical, randomly distributed holes marked from 1 to 9 in~\reffig{Fig1}. In our previous experiments these holes were shunted with 50~$\Omega$ loads to realize up to nine scattering channels. For the analysis of the fluctuation properties of the resonance frequencies we consider only $M=2$ scattering channels since, as was shown in Ref. \cite{Bialous2020}, for this case TIV is strongest in a given frequency interval. Furthermore, for increasing openness and internal absorption the overlap of the resonances becomes stronger and renders the identifiction of the resonance freqencies impossible. 

In order to measure the two-port scattering matrix $\hat{S}(\nu)$  two antennas of lengths 5.8~mm and pin diameter 0.9~mm were attached to the microwave cavity  at the positions marked by 1 and 2 in~\reffig{Fig1} and connected to an Agilent E8364B Vector Network Analyzer (VNA). The two antennas correspond to the $M=2$ scattering channels. A metallic perturber with perimeter $\approx 26$ cm and area $\approx 9$ cm$^2$ was placed inside the cavity and moved with a small external magnet marked by M$_P$ along the walls of the cavity in order to create different realizations of it. Cylindrical NiZn ferrites of diameter 33~mm and height 6~mm with saturation magnetization $2600$~Oe (manufactured by SAMWHA, South Korea) were inserted into the cavity and magnetized by two external NdFeB magnets of diameter 33~mm and height 30~mm and type N42 with coercity 11850~Oe (943~kA/m) below and above the cavity (marked by M$_1$ and M$_2$). The thereby generated homogenous magnetic field of strength $B\simeq 495$~mT induces a macroscopic magnetization of the ferrites across their cross sections. The precession of the magnetization around $B$ with the Larmor frequency $\omega_{o}=\gamma B$ and gyromagnetic ratio $\gamma$ = 32.2  GHz/T results in the appearance of ferromagnetic resonances at $\nu_{fr}$ = 15.9~GHz. Inducing TIV leads to different matrix elements $S_{12}(\nu)\ne S_{21}(\nu)$ of the measured scattering matrix, while for systems with preserved \Ti-invariance the scattering matrix is symmetric, $S_{12}(\nu)= S_{21}(\nu)$. Examples of measured spectra are presented in the upper right corner of~\reffig{Fig1}. The strength of TIV was controlled by varying the external magnetic field $B$ and depends on the frequency range. It is characterized by the parameter $\xi$. All measurements were done in the frequency range $\nu = 6 - 12$~GHz for $M=2$ scattering channels, and yielded values $\xi \simeq 0.19 - 0.49$. The size of $\xi$ was determined in~\cite{Bialous2020} by comparing the experimentally obtained cross-correlation coefficients 
\begin{equation}
\label{Eq.1}
C_{12}^{cross}=\frac{\Re[{\langle S^{fl}_{12}(\nu) S^{fl*}_{21}(\nu)\rangle]}}{\sqrt{\langle|(S^{fl}_{12}(\nu)|^2\rangle\langle|(S^{fl}_{21}(\nu)|^2\rangle}},
\end{equation}
where $S^{fl}_{12}(\nu)  = S_{12}(\nu) - \langle S_{12}(\nu)\rangle $ denotes the fluctuating part of the scattering matrix elements, with exact analytical results. Complete TIV yields a vanishing of the cross-correlation coefficient $C_{12}^{cross}=0$, because then $S^{fl}_{12}(\nu)$ and $S^{fl*}_{21}(\nu)$ are uncorrelated, and $C_{12}^{cross}=1$ for \Ti-invariant systems. The cross-correlation coefficient is particularly small in the frequency range 7.5~-~9.5~GHz~\cite{Bialous2020}, and strongest TIV $\xi = 0.49$ is achieved for $M=2$ open channels in the range 8~-~9~GHz. 

The transition between GOE and GUE was verified in~\cite{Bialous2020} based on an RMT approach for the fluctuation properties of the scattering matrix elements of chaotic scattering systems in the presence of absorption and partial \Ti-invariance violation. Here, we focus on the fluctuation properties in the resonance spectra. We restrict our analysis to $M = 2$ scattering channels as the strength $\xi$ of TIV is largest for this case in a given frequency range. Furthermore, with increasing number of scattering channels the overlap of the resonances increases so that the identification of the eigenfrequencies from their positions becomes more cumbersome. Actually, we are not able to identify complete sequences of levels, because of the strong overlap of resonances ascribed to large internal absorption in the cavity which is mainly caused by the lossy ferrites. The internal absorption $\gamma$ was determined in~\cite{Bialous2020} to $6<\gamma<15$ from the distributions of the diagonal elements of the measured scattering matrix. It is due to absorption of the electromagnetic field in the walls of the cavity and in the ferrites. The total absorption is given by $\gamma_{tot}=\gamma+\eta$ where $\eta=MT$ is the openness resulting from $M$ open channels with transmission coefficients $T$. It is commonly known that ferrite is a lossy material which, near the gyromagnetic resonance leads to an increase of the internal absorption and thus a degradation of the quality factor of the cavity. The absorption strength $\gamma$ was evaluated in 1~GHz  windows for $M=2$ open channels (antennas) in Ref.~\cite{Bialous2020}. With ferrites the absorption strength $\gamma$ is about 5 times larger than without them. Nevertheless, as we presented in~\cite{Bialous2016}, even a fraction of $\Phi=0.8$ identified levels suffices to classify a system with either GOE or GUE behavior according to its symmetry properties based on the analysis of missing level statistics. In the following section we address this approach and outline the procedure in detail for the present case of partially violated \Ti invariance.  
\section{Fluctuation properties of incomplete spectra in the presence of partial TIV\label{Missing}}
Before comparing the spectral properties of the ensemble of micwoave billiards with RMT predictions for universal quantum systems we removed system specific properties by unfolding the resonance frequencies $\nu_i$ to mean spacing unity~\cite{StoeckmannBuch2000,Haake2001} with Weyl's law, $\epsilon_i =N^{Weyl}(\nu_i)=\frac{A\pi}{c_0^2}\nu_i^2+\frac{\mathcal{L}}{2c}\nu_i+$ const., which applies to the smooth part of the integrated spectral density $N(\nu_i)=N^{Weyl}(\nu_i)+N^{fluc}(\nu_i)$. 

In order to get insight into short-range correlations in the eigenfrequency spectra of the microwave billiard we analyzed the distribution $P(s)$ of the spacings between adjacent eigenvalues $s_i=\epsilon_{i+1}-\epsilon_i$ and its cumulant $I(s)$, which has the advantage that it does not depend on the binning size of the histograms yielding $P(s)$. For the analysis of long-range correlations we considered the variance $\Sigma^2(L)=\left\langle\left( N(L)-\langle N(L)\rangle\right)^2\right\rangle$, of the number $N(L)$ of eigenvalues $\epsilon_i$ in an interval of length $L$, where $\langle N(L)\rangle =L$, and the rigidity $\Delta_3(L)=\left\langle \min_{a,b}\int_{e-L/2}^{e+L/2}\left[N(e)-a-be\right]^2de\right\rangle$ which provides a measure for the stiffness of a spectrum. Here, $\langle\cdot\rangle$ denotes the average over an ensemble of random matrices or over the eigenvalue spectra of different realizations of the microwave billiard. Both measures may be expressed in terms of the two-point cluster function $Y_2(r)$, i.e., the rescaled two-point correlation function of two eigenvalues at a distance $r$, as~\cite{Mehta2004}
\be
\Sigma^2(L)=L-2\int_0^L(L-r)Y_2(r)dr\, ,
\label{sigma2}
\ee
and
\be
\Delta_3(L)=\frac{L}{15}-\frac{1}{15L^4}\int_0^L(L-r)^3\left(2L^2-9rL-3r^2\right)Y_2(r)dr\, .
\label{delta3}
\ee
A further measure for long-range correlations is the power spectrum~\cite{Relano2002,Faleiro2004,Molina2007,Riser2017}
\begin{equation}
\label{PowerS}
S(\tau)=\left\langle\left\vert\frac{1}{\sqrt{N}}\sum_{q=0}^{N-1} \delta_q\exp\left(-\frac{2\pi i\tau q}{N}\right)\right\vert^2\right\rangle,
\end{equation}
with $\delta_q=\epsilon_{q+1}-\epsilon_1-q$ denoting the deviation of the $q$th nearest-neighbor spacing from its mean value $q$. In~\cite{Bialous2016,Dietz2017} the power spectrum was investigated for the incomplete spectra of microwave networks and billiards exhibiting either GOE or GUE statistics and it was shown that $\Sigma^2(L)$ and $S(\tau)$ are particularly sensitive to missing levels. The power spectrum $S(\tau)$ only depends on the ratio $\tilde\tau =\tau/N$ and exhibits for $\tilde\tau\ll 1$ a power law dependence $\langle S(\tilde\tau)\rangle\propto \tilde\tau^{-\alpha}$~\cite{Relano2002,Faleiro2004}, where for regular systems $\alpha =2$ and for chaotic ones $\alpha =1$ independently of whether \T invariance is preserved or not~\cite{Gomez2005,Salasnich2005,Santhanam2005,Relano2008,Faleiro2006}. It can be expressed in terms of the form factor $K(\tau)=1-b(\tau)$~\cite{Faleiro2004} where
\be
b(\tau)=\int_{-\infty}^\infty Y_2(r)e^{-ir\tau}dr
\label{form}
\ee
is the Fourier transform of the two-point cluster function. 

The spectral properties are compared to analytical expressions which were obtained for the two-point cluster functions and the nearest-neighbor spacing distribution for random matrices interpolating between GOE and GUE. The matrices are given in terms of a sum
\begin{equation}
H_{ij}=H_{\rm\mu\nu}^{(S)}+i\lambda H_{ij}^{(A)}.
\label{eqn:hamiltonian}
\end{equation}
of a real-symmetric random matrix $\hat H^{(S)}$ from the GOE and a real-antisymmetric one, $\hat H^{(A)}$. The matrix elements are uncorrelated Gaussian-distributed random numbers with zero mean and variance chosen equal to unity. The parameter $\xi$, which is related to $\lambda$ through $\lambda=\frac{\pi\xi}{\sqrt{N}}$, determines the magnitude of \T violation in units of the mean spacing. For $\xi =0$ $\hat H$ describes chaotic systems with preserved \T invariance, whereas for $\pi\xi /\sqrt{N}=1$ $\hat H$ is a random matrix from the GUE. However, the transition from GOE to GUE already takes place for $\xi\simeq 1$~\cite{Dietz2010}.

An exact analytical expression was derived for the nearest-neighbor spacing distribution in terms of a Taylor series in~\cite{Dietz1991,Mehta2004}. It was shown there that it is well approximated by the Wigner-like approximation derived in Ref.~\cite{Lenz1992} based on two-dimensional random matrices,
\be
P(s;\lambda)=s\sqrt{\frac{2+\lambda^2}{2}}c(\lambda)^2{\rm erf}\left(s\frac{c(\lambda)}{\lambda}\right)e^{-\frac{s^2c(\lambda)^2}{2}}
\label{ps}
\ee
with $\lambda=2\xi$,
\be 
c(\lambda)=\sqrt{\pi\frac{2+\lambda^2}{4}}\left[1-\frac{2}{\pi}\left(\tan^{-1}\left(\frac{\lambda}{\sqrt{2}}\right)-\frac{\sqrt{2}\lambda}{2+\lambda^2}\right)\right]
\ee
and ${\rm erf}(x)$ denoting the error function. Furthermore, in Ref.~\cite{Pandey1991,Bohigas1995} an analytical expression was derived for the two-point cluster function,
\begin{equation}
Y_2(L;\xi)=\det
\begin{pmatrix}
s(L) &-D(L;\xi)\\
-J(L;\xi) &s(L)
\end{pmatrix},
\label{yl}
\end{equation}
with~\cite{Pandey1991,Bohigas1995} 
\ba
s(L)&=&\frac{\sin\pi L}{\pi L},\\
D(L;\xi)&=&\frac{1}{\pi}\int_0^\pi{\rm d}xe^{2\xi^2 x^2}x\sin(Lx),\\
J(L;\xi)&=&\frac{1}{\pi}\int_\pi^\infty{\rm d}xe^{-2\xi^2x^2}\frac{\sin(Lx)}{x}.\label{jl}
\ea
Expressions for $\Sigma^2(L;\xi)$ and $\Delta_3(L;\xi)$ for the case of partial TIV are obtained by inserting $Y_2(L;\xi)$ into Eqs.~(\ref{sigma2}),~(\ref{delta3}) and~(\ref{form}).

\begin{figure}[h!]
\includegraphics[width=\linewidth]{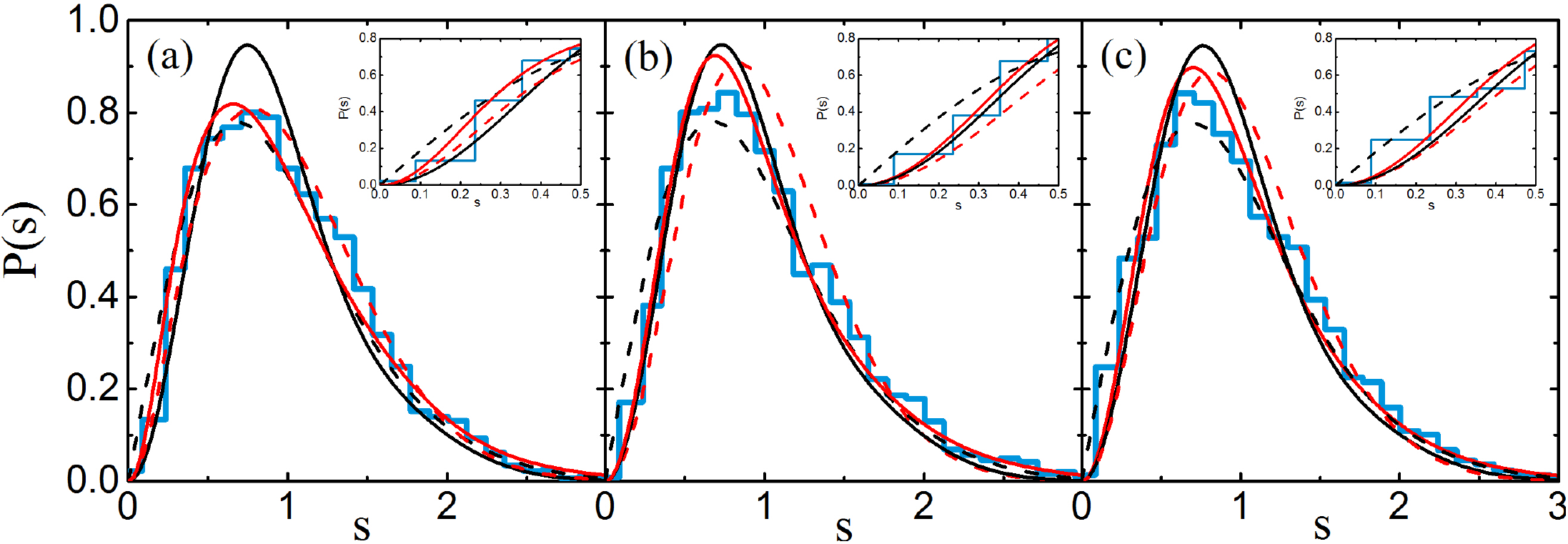}
	\caption{Nearest-neighbor spacing distributions in the frequency ranges 6.5 - 8 GHz (a), 8 - 9 GHz (b), and 9.2 - 11.5 GHz (c) corresponding to a strength of \Ti-invariance violation $\xi=0.19,\, 0.35,\,0.49$, respectively. The turquoise histograms show the experimental results, the red dashed lines the RMT curve for the intermediate case between GOE and GUE. The solid red line shows the result for the intermediate case with a fraction of $\Phi=0.83,\, 0.81,\, 0.85$ identified eigenfrequencies, respectively. The black-dashed and, black solid lines show the corresponding results for GOE and GUE for these values of $\Phi$, respectively. 
}
\label{fig2}
\end{figure}

\begin{figure}[h!] 
\includegraphics[width=\linewidth]{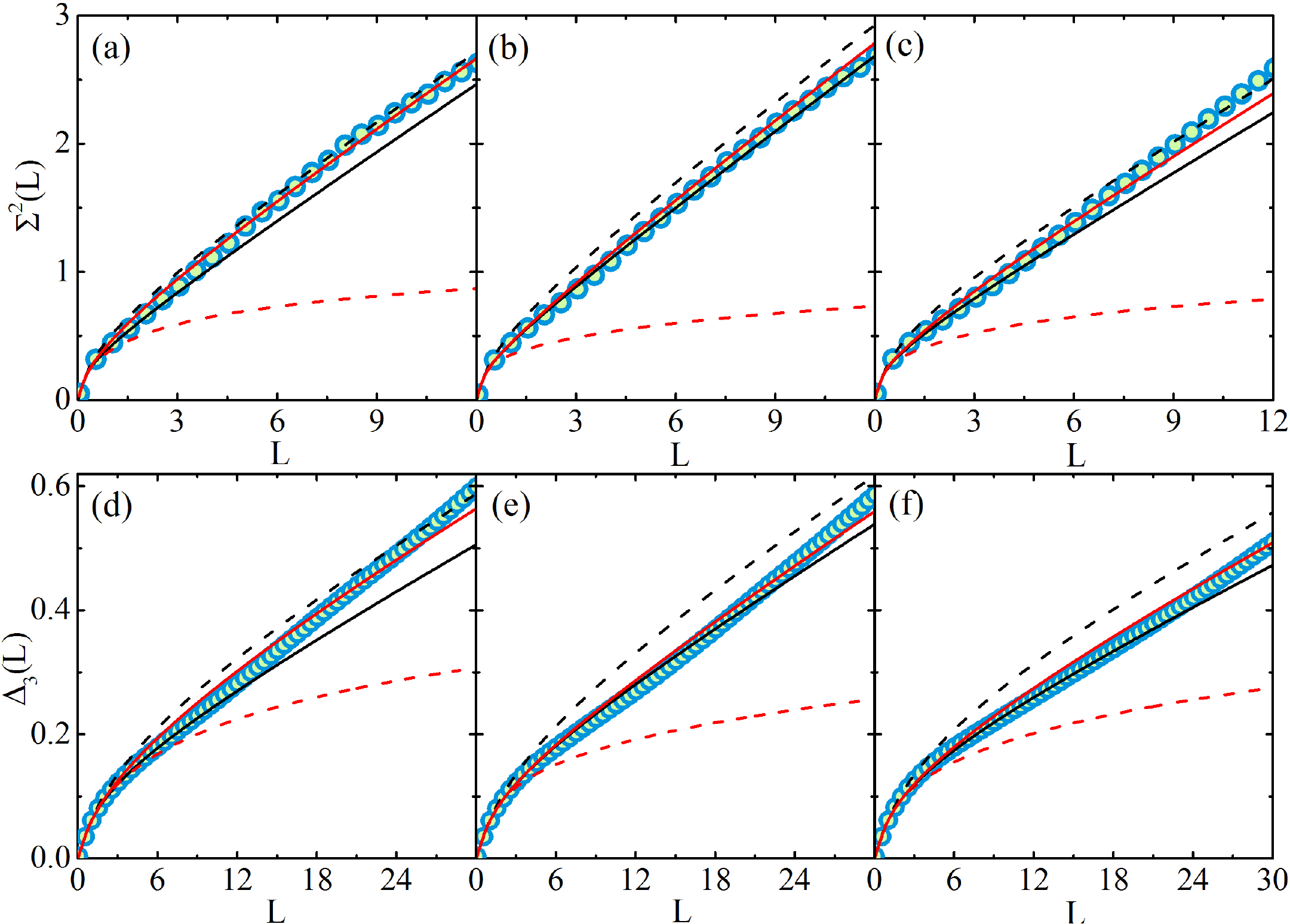}
\caption{Same as~\reffig{fig2} for the number variance [(a)-(c)] and the rigidity [(d)-(f)]. The experimental results are shown as turquoise circles.
}
\label{fig3}
\end{figure}

\begin{figure}[h!]
\includegraphics[width=\linewidth]{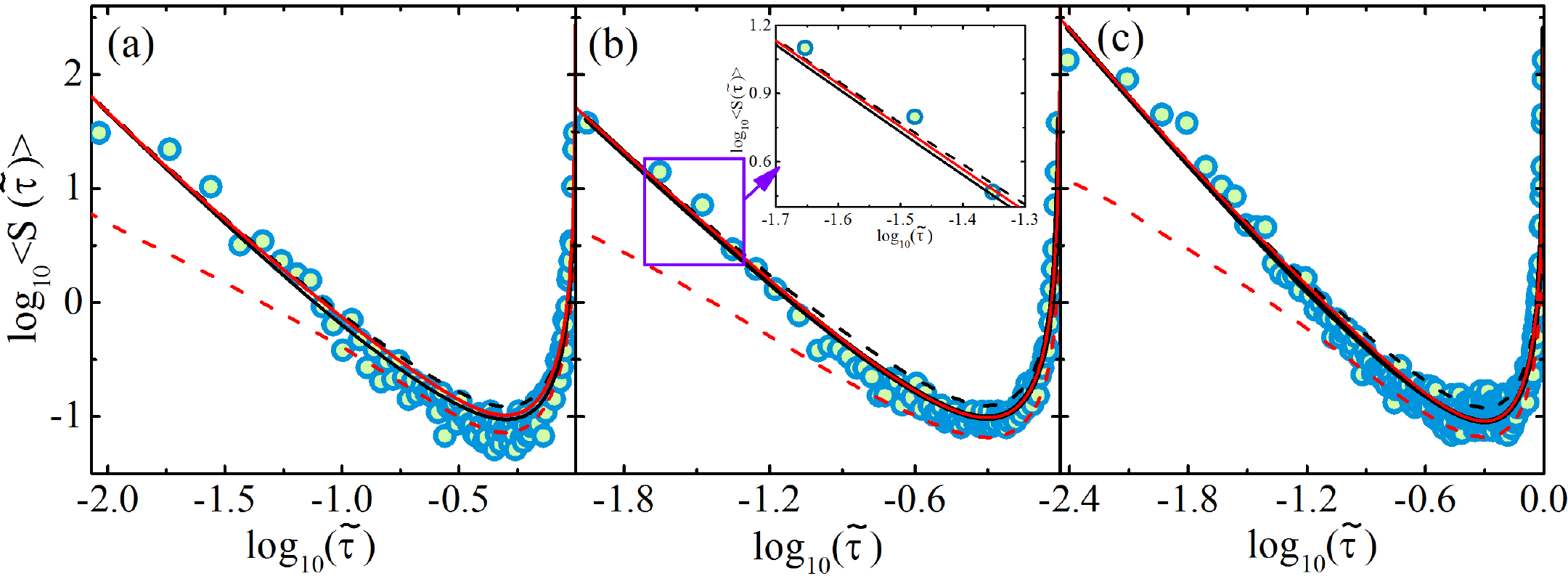}
\caption{Same as~\reffig{fig2} for the power spectrum. The experimental results are shown as turquoise circles. 
}
\label{fig4}
\end{figure}

We analyzed the spectral properties of the microwave billiards in three frequency ranges, 6.5~-~8~GHz, 8~-~9~GHz and 9.2~-~11.5 GHz, since the ferrite properties depend strongly on the microwave frequency and, accordingly, the degree of \Ti-invariance violation varies with frequency. For each of these three ranges we randomly selected  25 realizations of the cavity, yielding 110, 90, and 258 eigenfrequencies for each eigenfrequency sequence, respectively. The results for the nearest-neighbor spacing distributions, the number variance, the rigidity and the power spectrum are shown in Figs.~\ref{fig2}~-~\ref{fig4}. Clear deviations from the corresponding theoretical curves (red dash-dotted lines) are visible. 

\begin{figure}[!th]
\includegraphics[width=0.6\linewidth]{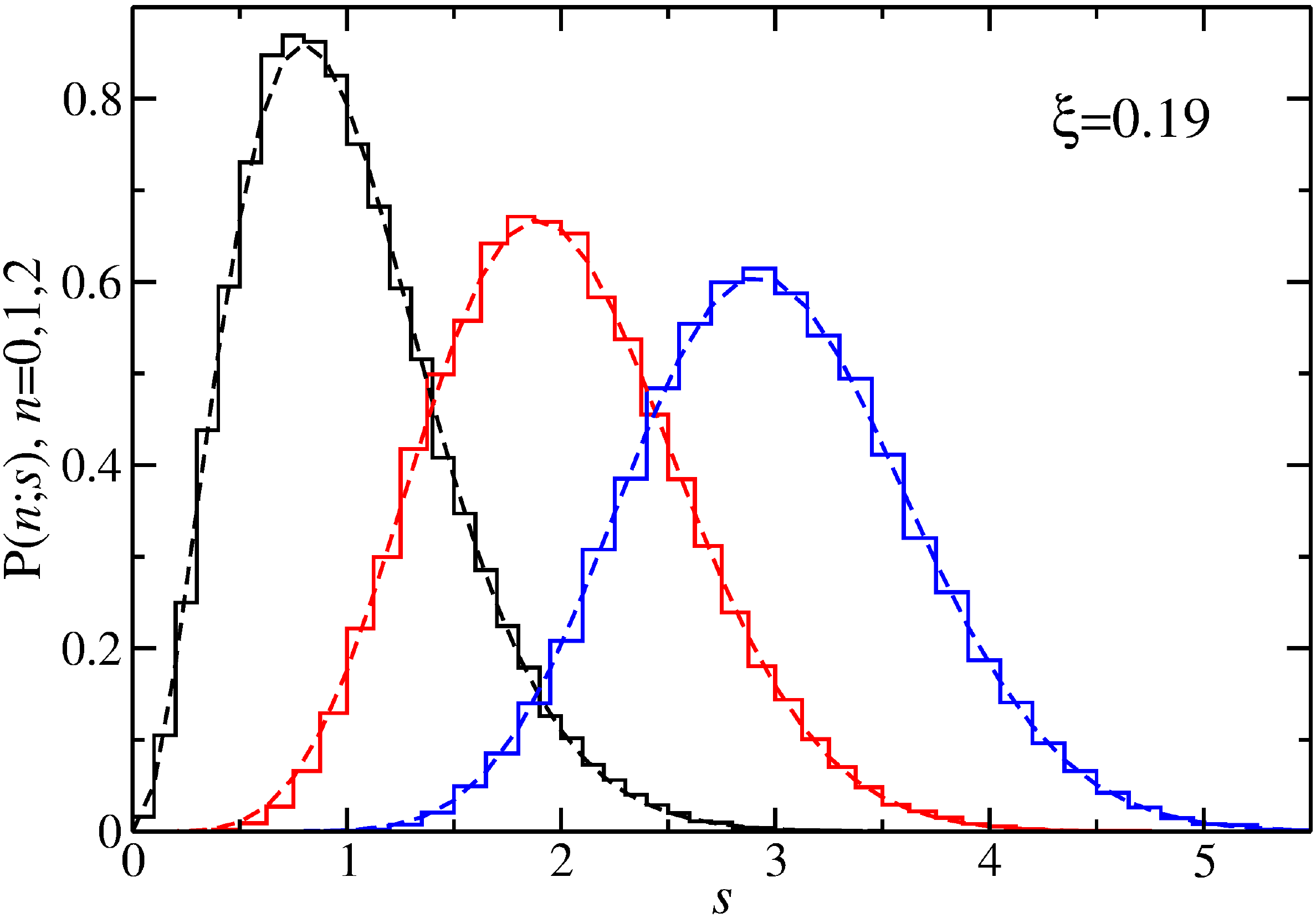}
\includegraphics[width=0.6\linewidth]{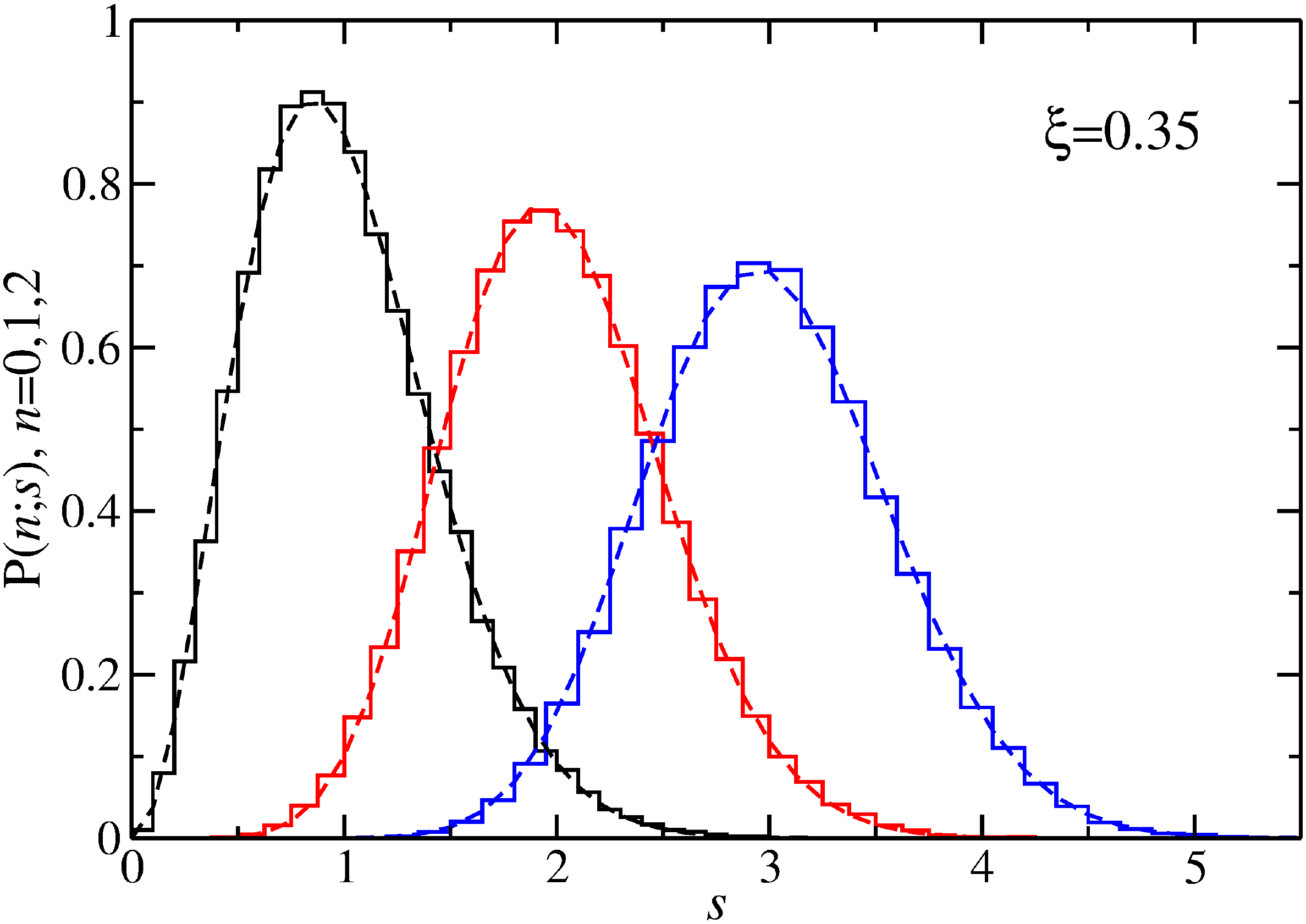}
\includegraphics[width=0.6\linewidth]{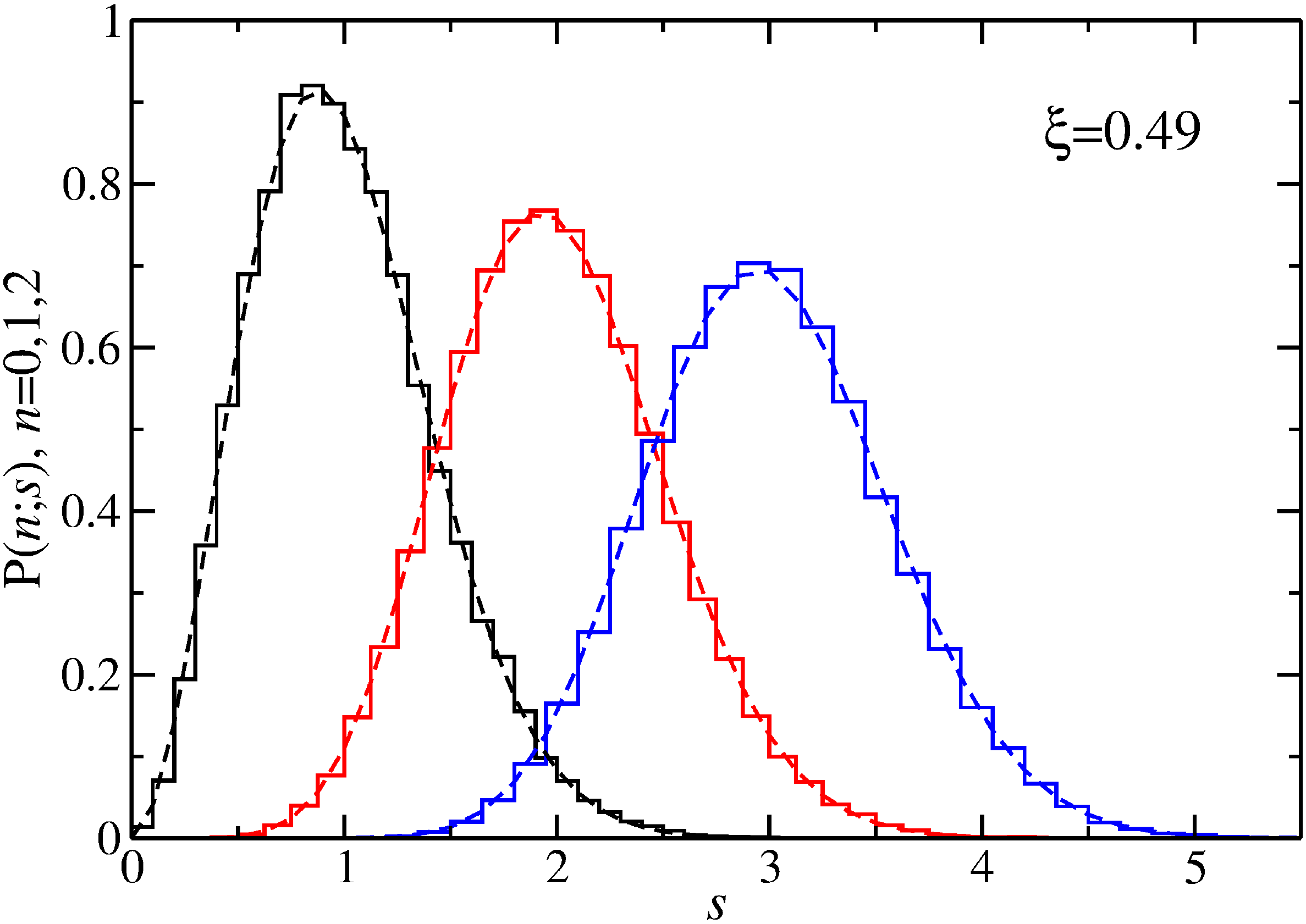}
\caption{Nearest-neighbor (black histogram), next-nearest neighbor (red histogram) and second-nearest neighbor (blue histogram) spacing distributions for partial \Ti-invariance violation where the size $\xi$ is indicated in the panels. The dashed curves were obtained from a fit of $\tilde P(s)=\gamma s^\mu e^{-\chi s^2}$ to the histogram of corresponding color~\cite{Bohigas2006,Stoffregen1995}.}
\label{Fig2}
\end{figure}

The discrepancies are attributed to missing levels. In order to derive RMT predictions for the fluctuation properties in incomplete spectra of quantum systems experiencing a partial TIV, we proceeded as in Ref.~\cite{Bohigas2004} on the basis of the analytical results Eqs.~(\ref{ps}) -~(\ref{jl}). The derivation relies on the assumption that a fraction $1-\Phi$ of levels are missed randomly. The nearest-neighbor spacing distribution is expressed in terms of the $(n+1)$st nearest-neighbor spacing distribution $P(n,s)$ with $P(0,s)=P(s)$ of the corresponding complete spectrum,
\begin{align}
\label{abst}
p(s)&=\sum_{n=0}^{M}(1-\Phi)^nP\left(n;\frac{s}{\Phi}\right)\\
&\simeq\sum_{n=0}^{K-1}(1-\Phi)^nP\left(n;\frac{s}{\Phi}\right)\nonumber\\
&+\sum_{n=K}^{M}\frac{(1-\Phi)^n}{\sqrt{2\pi V^2(n)}}\exp\left(-\frac{1}{2V^2(n)}\left[\frac{s}{\Phi}-n-1\right]^2\right),\nonumber
\end{align}
where
\be
V^2(n)\simeq \Sigma^2(L=n)-\frac{1}{6}.
\ee
We chose $K=3$ and obtained $P(n;s)$ with $n=1,2$ by computing the ensemble averages of the normalized next and 2nd-next nearest-neighbor spacing distributions of 500 $500\times 500$ random matrices of the form~\refeq{eqn:hamiltonian} and fitting $\tilde P(s)=\gamma s^\mu e^{-\chi s^2}$ to the resulting distributions, as illustrated in~\reffig{Fig2}.
We checked that for larger values of $n$ $P(n;s)$ is well approximated by a Gaussian with variance $V^2(n)$ centered at $n+1$~\cite{Bohigas1999}.

Since we assume that levels are missing randomly, the $n$-point correlation functions keep their form when a fraction $(1-\Phi)$ of the levels is extracted and the remaining ones are rescaled with $\Phi$~\cite{Bohigas2004},
\be
y_2(r)=Y_2\left(\frac{r}{\Phi}\right).
\ee
Using this feature of the two-point cluster function and Eqs.~(\ref{sigma2}),~(\ref{delta3}) and~(\ref{form}) yields~\cite{Bohigas2004}
\begin{equation}
\sigma^2(L)=(1-\Phi)L+\Phi^2\Sigma^2\left(\frac{L}{\Phi}\right),
\label{sigma2v}
\end{equation}
\begin{equation}
\delta_3(L)=(1-\Phi)\frac{L}{15}+\Phi^2\Delta_3\left(\frac{L}{\Phi}\right).
\label{delta3v}
\end{equation}
and
\begin{eqnarray}
s(\tilde\tau) &=&\nonumber
\frac{\Phi}{4\pi^2}\left[\frac{K\left(\Phi\tilde\tau\right)-1}{\tilde\tau^2}+\frac{K\left(\Phi\left(1-\tilde\tau\right)\right)-1}{(1-\tilde\tau)^2}\right]\\
&+& \frac{1}{4\sin^2(\pi\tilde\tau)} -\frac{\Phi^2}{12},
\label{noisev}
\end{eqnarray}
where $\vert\tilde\tau\vert\leq 1$. In that range of $\tilde\tau$ $b(\tilde\tau)=1-\vert\tilde\tau\vert$ for the GUE and $b(\tilde\tau)=1-2\vert\tilde\tau\vert+\vert\tilde\tau\vert\ln\left(1+2\vert\tau\vert\right)$ for the GOE.

\begin{figure}[h!]
\includegraphics[width=0.9\linewidth]{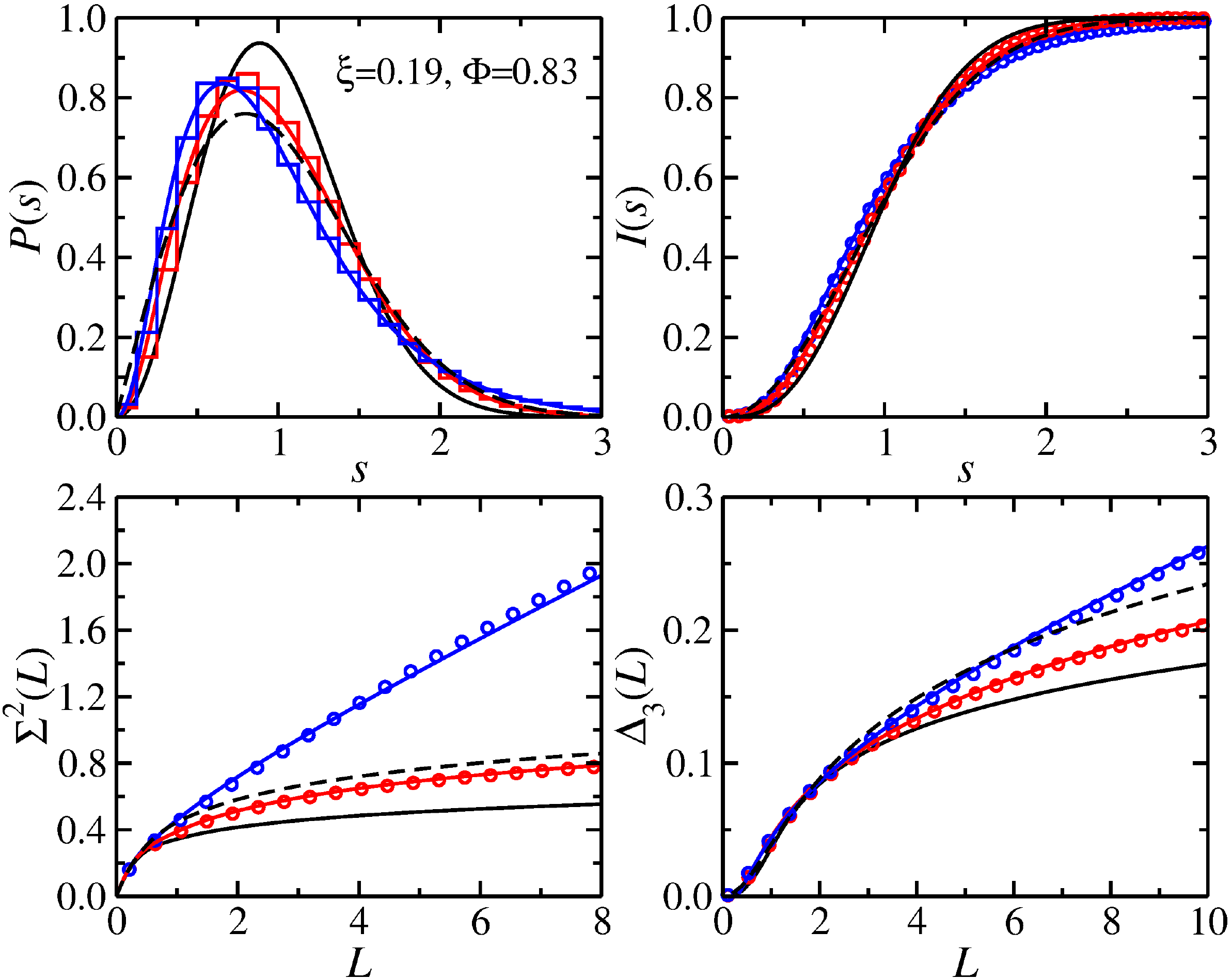}
\includegraphics[width=0.9\linewidth]{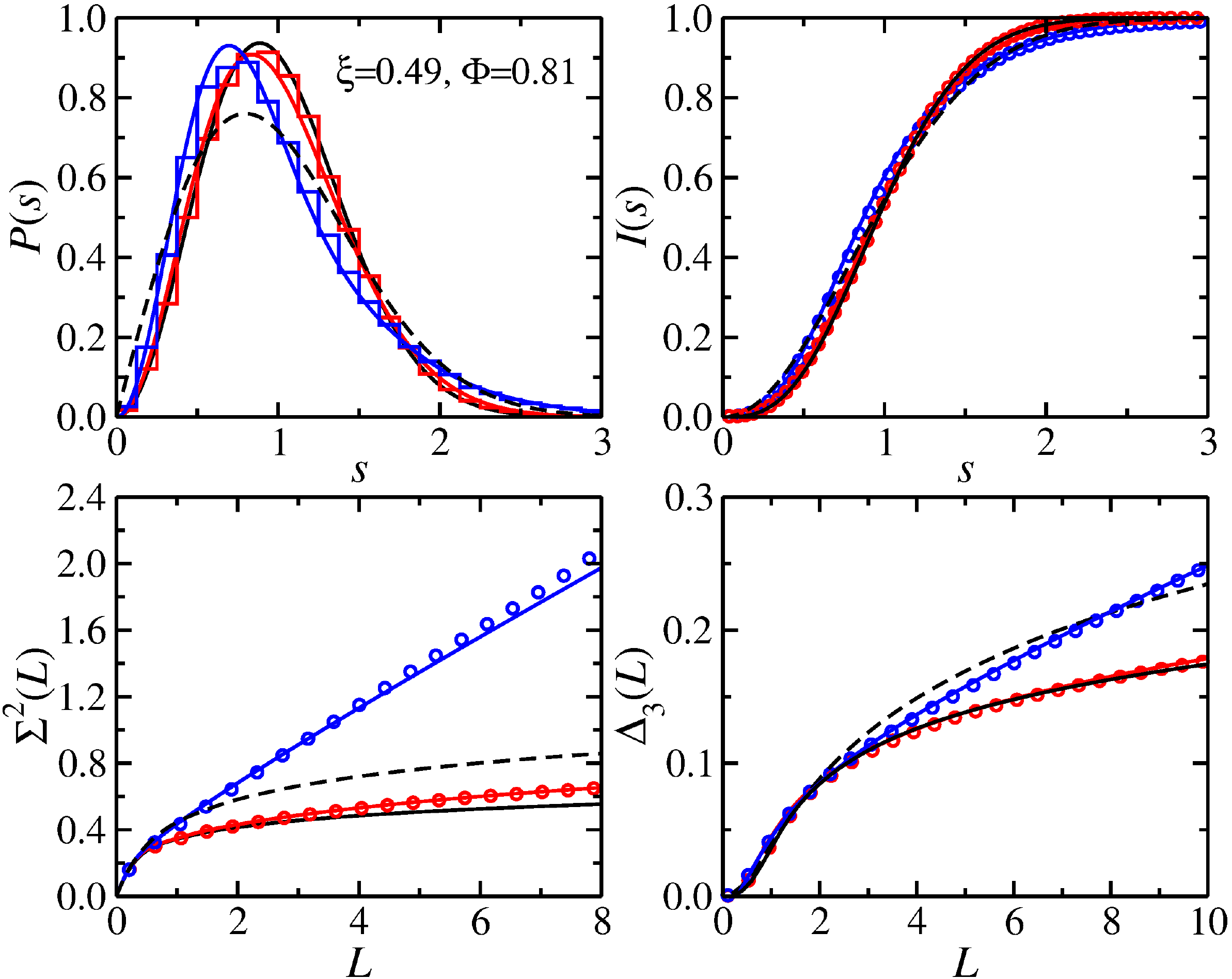}
\includegraphics[width=0.9\linewidth]{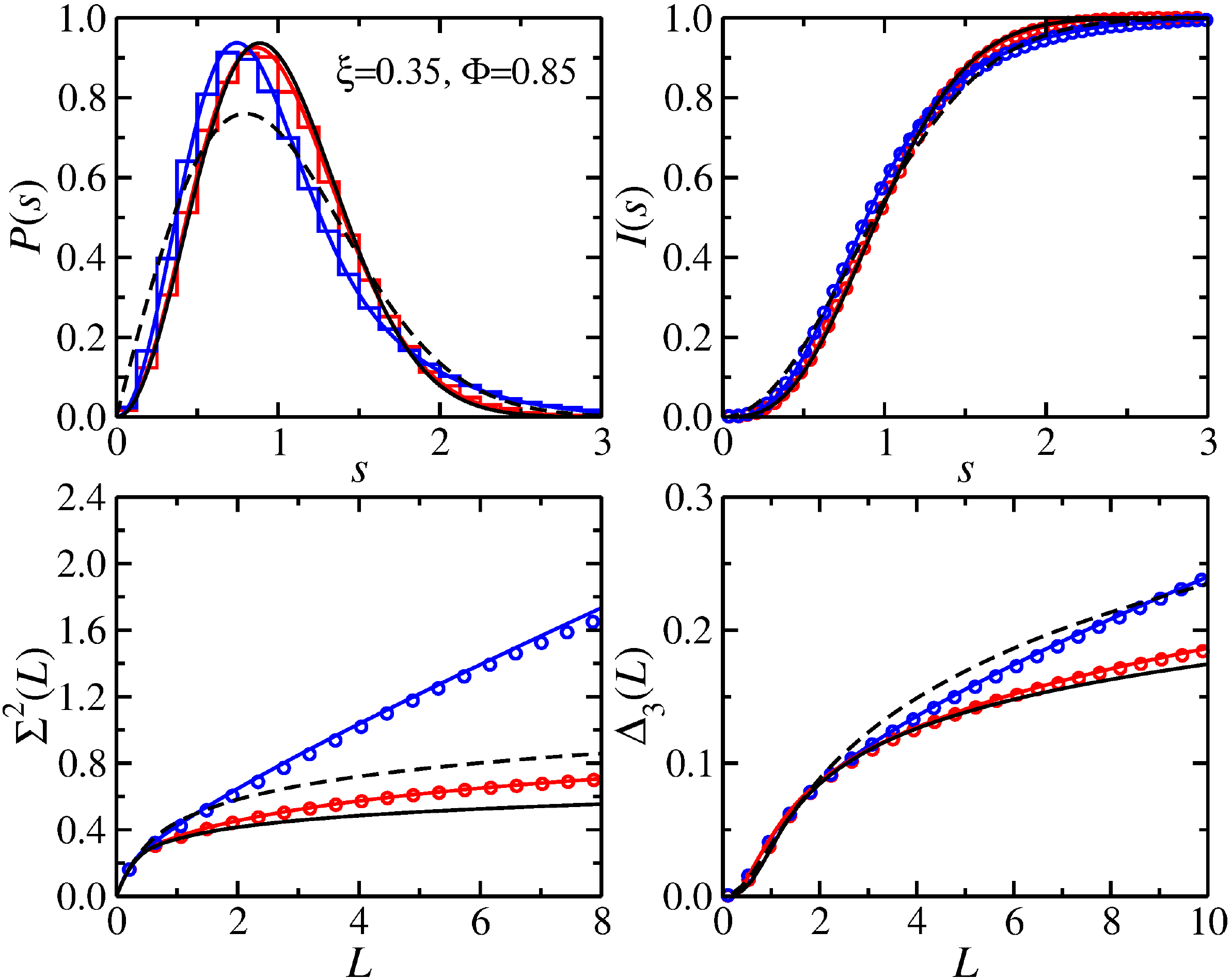}
\caption{Comparison of the anlytical results (solid lines) for $\Phi =1$ (red) and for the values of $\Phi$ given in the panels (blue) with RMT simulations (histograms and circles) which were performed based on the random matrix given in~\refeq{eqn:hamiltonian}.
}
\label{fig5}
\end{figure}

\begin{figure}[h!]
\includegraphics[width=0.9\linewidth]{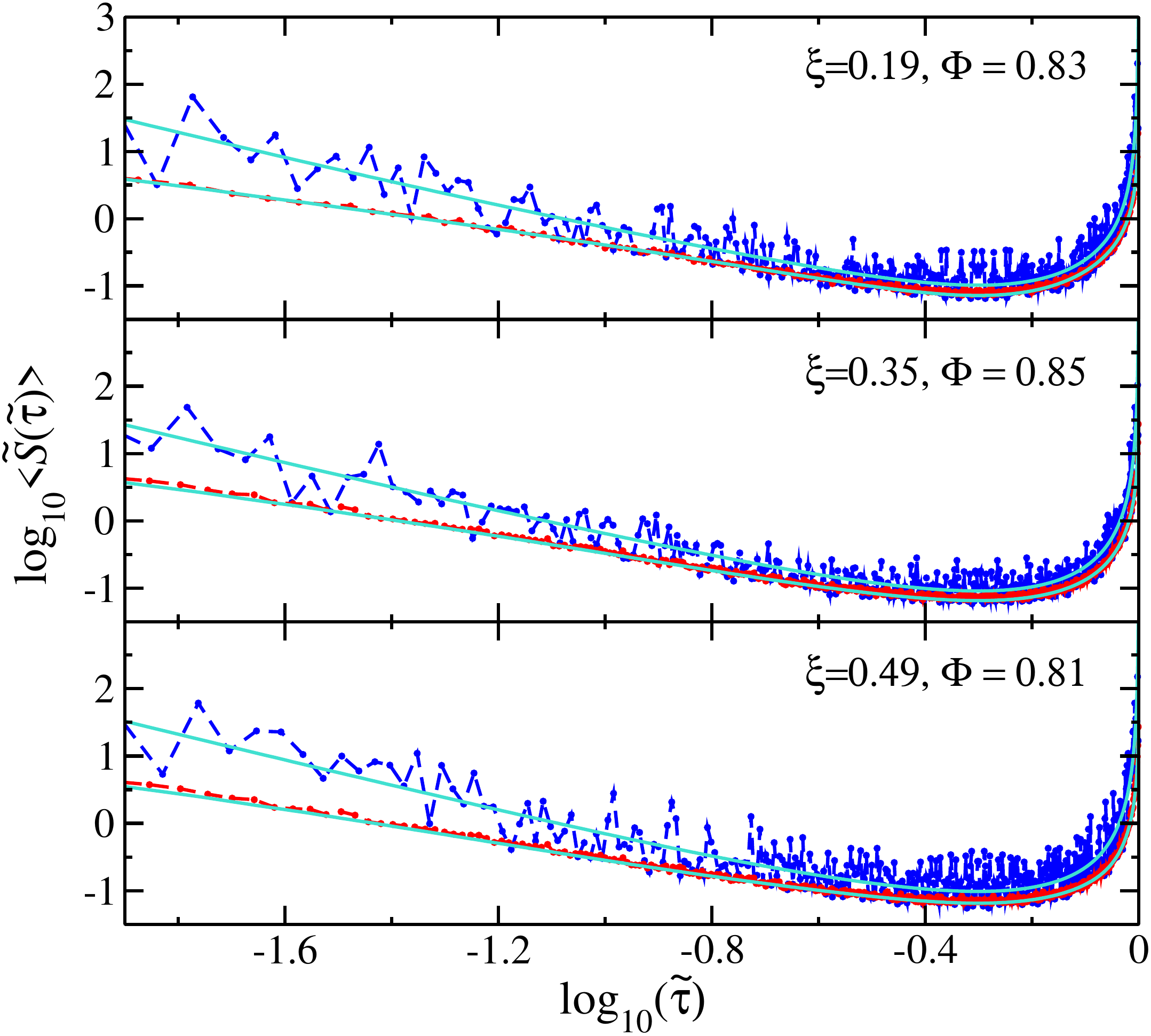}
\caption{Comparison of the power spectra obtained from RMT simulations based on~\refeq{eqn:hamiltonian} for complete spectra (red dots) and incomplete ones (blue dots) for partial TIV in comparison to the corresponding analytical results (turquoise solid lines) obtained from~\refeq{noisev}. The values of the fraction $\Phi$ and the strength of  TIV $\xi$ are given in the panels.}
\label{fig5b}
\end{figure}

Before comparing these results to experimental ones we validated them with RMT simulations obtained for ensembles of 300 $700\times 700$-dimensional random matrices $\hat H$ as given in~\refeq{eqn:hamiltonian}, where for illustration, we chose in~\reffig{fig5} and in ~\reffig{fig5b} for $\xi$ and $\Phi$ the values obtained from the analysis of the experimental results. The analytical results for $\Phi =1$ and the values of $\Phi$ given in the figures are shown as red and blue solid lines, respectively, those obtained from RMT simulations as histograms and circles of corresponding color. The black solid and dashed lines show the GUE and GOE curves, respectively. The good agreement between the analytical curves and those obtained from the RMT simulations corroborates the applicability of our RMT approach to the incomplete spectra of systems with partial TIV.

\begin{figure}[h!]
\includegraphics[width=0.9\linewidth]{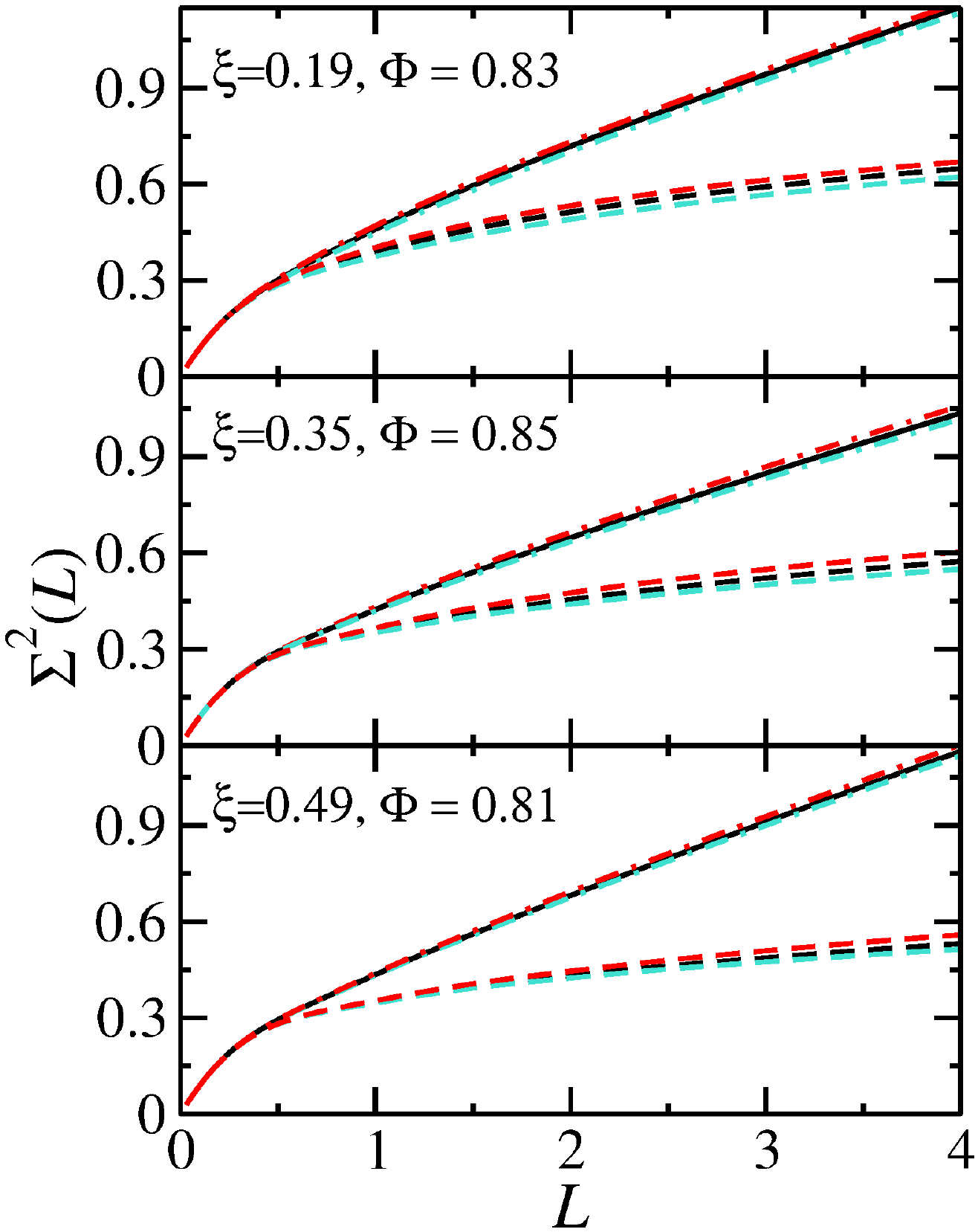}
	\caption{Comparison of $\Sigma^2(L)$ obtained from~\refeq{sigma2v} for $\Phi =0$ (black dashed line) and incomplete (black full line) spectra for the values of $\xi$ and $\Phi$ given in the panels. They are compared to the corresponding analytical results with 20$\%$ added to (red lines) and subtracted (turquoise lines) from the corresponding value of $\xi$.}
\label{fig6}
\end{figure}

The statistical measures depend on two parameters, the strength of TIV $\xi$ and the fraction of identified levels ${\Phi}$. The values of $\xi$ were determined in 1~GHz windows in~\cite{Bialous2020} by fitting exact analytical expressions for the cross-correlation coefficients to the experimental results and refined by comparing the distributions of the off-diagonal elements of the measured scattering matrix and the experimentally determined enhancement factors to RMT predictions. To obtain the fraction of missing levels we used the power spectrum $\langle s(\tilde\tau)\rangle$ ~\refeq{noisev}. This measure is particularly sensitive to changes in the value of $\Phi$~\cite{Bialous2016,Dietz2017}, and above all its asymptotic behavior does not depend on the universality class. This is illustrated in the inset of~\reffig{fig4} (b). The curves for the GOE (black dashed line), the GUE (black solid line) and the intermediate case between GOE and GUE (red solid line) lie on top of each other. We inserted the resulting values of $\xi$ into the expression~\refeq{noisev} and fitted then $\langle s(\tilde\tau)\rangle$ in the asypmtotic region to the experimental curves, yielding that in the frequency ranges 6.5~-~8~GHz, 8~-~9~GHz, and 9.2~-~11.5~GHz, respectively, $83 \pm 3\%$, $81\pm 3 \%$, and $85\pm 3 \%$ of the resonances were identified. The errors comprise those resulting from the fitting procedure and the ensemble averaging. Note that billiard systems have the great advantage over, e.g. nuclear systems, that an analytical expression, namely the Weyl formula, exists for the average integrated spectral density thus providing an estimate for the value of $\Phi$ and the accuracy of the obtained value, so that a procedure like the one described in~\cite{Casal2021} is not needed. The values for $\Phi$ indeed agree with those deduced from Weyl's law within the error. The number variance $\Sigma^2(L)$ depends on the universality class and thus is used to confirm the value of $\xi$ obtained from the cross-correlation coefficient. Yet, as illustrated in~\reffig{fig6} $\Sigma^2(L)$ barely changes when varying $\xi$ by less than $\approx 20\%$ of its value, so that either an ensemble of high statistical relevance is needed or an additional measure like the cross-correlation coefficient to obtain a statistically relevant estimate. 
Shown are the results deduced from~\refeq{sigma2v} for $\Sigma^2(L)$ for complete (black dashed line) and incomplete (black full line) spectra at the values of $\xi$ determined from the cross-correlation coefficient. They are compared to the curves after subtracting (turquoise lines) and adding (red lines) 20~$\%$ of the values of $\xi$ indicated in the panels. The rigidity $\Delta_3(L)$ is given as an integral over $\Sigma^2(L)$ and thus corresponds to a smoothing of it. It, in fact, leads to an enhancement of the differences between the curves corresponding to different values of $\xi$. The resulting curves are shown as red solid lines in Figs.~\ref{fig2}~-~\ref{fig4}. The agreement between the experimental and theoretical curves is good for the long-range correlation measures, whereas slight deviations are observed for the nearest-neighbor spacing distribution, as illustrated in the insets of~\reffig{fig2}. These may be attributed to the fact that eigenfrequencies may not be detected if two resonances are overlapping, that is, when they are too close to each other. This is reflected in the deviations of the experimental nearest-neighbor spacing distributions from the RMT predictions observed at small spacings. Note, that the probability of a close encounter of two eigenfrequencies is small in chaotic systems, as reflected in the nearest-neighbor spacing distribution, which vanishes for spacing zero. Thus, randomness of missing levels is ensured by choosing randomly 25 different realizations of the cavity, except for small spacings. However, in experiments with microwave billiards and networks the realization of ensembles of level sequences might not be possible, so that one has to cope with randomly missing levels and also a large fraction of systematically missing ones. Such a situation indeed was encountered in an experiment with a microwave cavity~\cite{Lawniczak2018} due to the large spectral density which led to level clustering. Yet, this can be incorporated into the appropriate random matrix model by extracting from the list of eigenvalues those with spacings to surrounding ones below a certain threshold. 

\section{Conclusions\label{Concl}}
We investigated the fluctuations in the resonance frequency spectra of microwave billiards with the shape of a fully chaotic quarter bowtie microwave billiard subject to partial TIV of varying strength. The microwave billiards simulate the properties of a two-dimensional quantum billiard with $M=2$ equivalent open channels subject to TIV. The strength of TIV is characterized by a parameter $\xi$ and the number of missing levels is expressed by the fraction $\Phi$ of the complete sequence of levels which could be identified. We analyze such a situation based on an RMT approach introduced in~\cite{Bohigas2006} to derive expressions for statistical measures of short- and long-range correlations in the spectra allowing to unambiguously assign the strength of TIV $\xi$ and the fraction of missing levels $1-\Phi$. Since the power spectrum $\langle s(\tilde\tau)\rangle$ depends for small values of $\tilde\tau\ll 1$ only weakly on the universality class and thus on the value of $\xi$ and, on the other hand is sensitive to the fraction $\Phi$ of identified levels, it provides a suitable measure for the determination of $\Phi$. Then, the number variance $\sigma^{2}(L)$ allows to refine or confirm the value of $\xi$ determined from the cross-correlation coefficient. The RMT approach presented in this article reproduces very well the results obtained from RMT simulations and the experimental results for all available sets of parameters $\Phi,\, \xi$, thus corroborating that it may serve as a tool to determine the extend of incompleteness of a spectrum and the value of $\xi$ for dissipative quantum systems with classically chaotic counterpart subject to partial TIV. 

The case of randomly and systematically missing levels has been addressed, e.g., in Ref.~\cite{Che2021}. There quantum graphs and microwave networks were investigated which exhibit localized states that can be identified and extracted from the level dynamics generated by varying a parameter. This yields incomplete level sequences with systematically missing levels in addition to due to experimental reasons randomly missing ones. In these systems the effect of the non-universal states on the spectral dynamics is only visible for distances below 2-3 mean level spacings and does not impede the determination of the fraction of missing levels and of the universality class.   

\section{Acknowledgement}
This work was supported in part by the National Science Center, Poland, Grant No. UMO-2018/30/Q/ST2/00324. B.D. thanks the National Natural Science Foundation of China for financial support through Grants Nos. 11775100, 11961131009 and 12047501. Supported by the 111 Project under Grant No. B20063. 
\bibliography{ref_spaghetti,ref_triv,References}

\begin{thebibliography}{75}
\expandafter\ifx\csname natexlab\endcsname\relax\def\natexlab#1{#1}\fi
\expandafter\ifx\csname bibnamefont\endcsname\relax
  \def\bibnamefont#1{#1}\fi
\expandafter\ifx\csname bibfnamefont\endcsname\relax
  \def\bibfnamefont#1{#1}\fi
\expandafter\ifx\csname citenamefont\endcsname\relax
  \def\citenamefont#1{#1}\fi
\expandafter\ifx\csname url\endcsname\relax
  \def\url#1{\texttt{#1}}\fi
\expandafter\ifx\csname urlprefix\endcsname\relax\def\urlprefix{URL }\fi
\providecommand{\bibinfo}[2]{#2}
\providecommand{\eprint}[2][]{\url{#2}}

\bibitem[{\citenamefont{Berry}(1979)}]{Berry1979}
\bibinfo{author}{\bibfnamefont{M.}~\bibnamefont{Berry}},
  \emph{\bibinfo{title}{Structural Stability in Physics}}
  (\bibinfo{address}{Berlin}, \bibinfo{year}{1979}).

\bibitem[{\citenamefont{Casati et~al.}(1980)\citenamefont{Casati, Valz-Gris,
  and Guarnieri}}]{Casati1980}
\bibinfo{author}{\bibfnamefont{G.}~\bibnamefont{Casati}},
  \bibinfo{author}{\bibfnamefont{F.}~\bibnamefont{Valz-Gris}},
  \bibnamefont{and}
  \bibinfo{author}{\bibfnamefont{I.}~\bibnamefont{Guarnieri}},
  \bibinfo{journal}{Lett. Nuovo Cimento} \textbf{\bibinfo{volume}{28}},
  \bibinfo{pages}{279} (\bibinfo{year}{1980}).

\bibitem[{\citenamefont{Bohigas et~al.}(1984)\citenamefont{Bohigas, Giannoni,
  and Schmit}}]{Bohigas1984}
\bibinfo{author}{\bibfnamefont{O.}~\bibnamefont{Bohigas}},
  \bibinfo{author}{\bibfnamefont{M.~J.} \bibnamefont{Giannoni}},
  \bibnamefont{and} \bibinfo{author}{\bibfnamefont{C.}~\bibnamefont{Schmit}},
  \bibinfo{journal}{Phys. Rev. Lett.} \textbf{\bibinfo{volume}{52}},
  \bibinfo{pages}{1} (\bibinfo{year}{1984}).

\bibitem[{\citenamefont{Mehta}(2004)}]{Mehta2004}
\bibinfo{author}{\bibfnamefont{M.}~\bibnamefont{Mehta}},
  \emph{\bibinfo{title}{Random Matrices}} (\bibinfo{publisher}{Elsevier
  Science}, \bibinfo{year}{2004}).

\bibitem[{\citenamefont{Vina et~al.}(1998)\citenamefont{Vina, Potemski, and
  Wang}}]{Vina1998}
\bibinfo{author}{\bibfnamefont{L.}~\bibnamefont{Vina}},
  \bibinfo{author}{\bibfnamefont{M.}~\bibnamefont{Potemski}}, \bibnamefont{and}
  \bibinfo{author}{\bibfnamefont{W.~I.} \bibnamefont{Wang}},
  \bibinfo{journal}{Phys.-Usp.} \textbf{\bibinfo{volume}{41}},
  \bibinfo{pages}{153} (\bibinfo{year}{1998}).

\bibitem[{\citenamefont{Zimmermann et~al.}(1988)\citenamefont{Zimmermann,
  K\"oppel, Cederbaum, Persch, and Demtr\"oder}}]{Zimmermann1988}
\bibinfo{author}{\bibfnamefont{T.}~\bibnamefont{Zimmermann}},
  \bibinfo{author}{\bibfnamefont{H.}~\bibnamefont{K\"oppel}},
  \bibinfo{author}{\bibfnamefont{L.~S.} \bibnamefont{Cederbaum}},
  \bibinfo{author}{\bibfnamefont{G.}~\bibnamefont{Persch}}, \bibnamefont{and}
  \bibinfo{author}{\bibfnamefont{W.}~\bibnamefont{Demtr\"oder}},
  \bibinfo{journal}{Phys. Rev. Lett.} \textbf{\bibinfo{volume}{61}},
  \bibinfo{pages}{3} (\bibinfo{year}{1988}).

\bibitem[{\citenamefont{Sirko et~al.}(1993)\citenamefont{Sirko, Bellermann,
  Haffmans, Koch, and Richards}}]{Sirko1993}
\bibinfo{author}{\bibfnamefont{L.}~\bibnamefont{Sirko}},
  \bibinfo{author}{\bibfnamefont{M.~R.~W.} \bibnamefont{Bellermann}},
  \bibinfo{author}{\bibfnamefont{A.}~\bibnamefont{Haffmans}},
  \bibinfo{author}{\bibfnamefont{P.~M.} \bibnamefont{Koch}}, \bibnamefont{and}
  \bibinfo{author}{\bibfnamefont{D.}~\bibnamefont{Richards}},
  \bibinfo{journal}{Phys. Rev. Lett.} \textbf{\bibinfo{volume}{71}},
  \bibinfo{pages}{2895} (\bibinfo{year}{1993}).

\bibitem[{\citenamefont{Sirko and Koch}(2002)}]{Sirko2002}
\bibinfo{author}{\bibfnamefont{L.}~\bibnamefont{Sirko}} \bibnamefont{and}
  \bibinfo{author}{\bibfnamefont{P.~M.} \bibnamefont{Koch}},
  \bibinfo{journal}{Phys. Rev. Lett.} \textbf{\bibinfo{volume}{89}},
  \bibinfo{pages}{274101} (\bibinfo{year}{2002}).

\bibitem[{\citenamefont{St{\"o}ckmann and Stein}(1990)}]{Stoeckmann1990}
\bibinfo{author}{\bibfnamefont{H.-J.} \bibnamefont{St{\"o}ckmann}}
  \bibnamefont{and} \bibinfo{author}{\bibfnamefont{J.}~\bibnamefont{Stein}},
  \bibinfo{journal}{Phys. Rev. Lett.} \textbf{\bibinfo{volume}{64}},
  \bibinfo{pages}{2215} (\bibinfo{year}{1990}).

\bibitem[{\citenamefont{Gr\"af et~al.}(1992)\citenamefont{Gr\"af, Harney,
  Lengeler, Lewenkopf, Rangacharyulu, Richter, Schardt, and
  Weidenm\"uller}}]{Graef1992}
\bibinfo{author}{\bibfnamefont{H.-D.} \bibnamefont{Gr\"af}},
  \bibinfo{author}{\bibfnamefont{H.~L.} \bibnamefont{Harney}},
  \bibinfo{author}{\bibfnamefont{H.}~\bibnamefont{Lengeler}},
  \bibinfo{author}{\bibfnamefont{C.~H.} \bibnamefont{Lewenkopf}},
  \bibinfo{author}{\bibfnamefont{C.}~\bibnamefont{Rangacharyulu}},
  \bibinfo{author}{\bibfnamefont{A.}~\bibnamefont{Richter}},
  \bibinfo{author}{\bibfnamefont{P.}~\bibnamefont{Schardt}}, \bibnamefont{and}
  \bibinfo{author}{\bibfnamefont{H.~A.} \bibnamefont{Weidenm\"uller}},
  \bibinfo{journal}{Phys. Rev. Lett.} \textbf{\bibinfo{volume}{69}},
  \bibinfo{pages}{1296} (\bibinfo{year}{1992}).

\bibitem[{\citenamefont{Sridhar and Kudrolli}(1994)}]{Sridhar1994}
\bibinfo{author}{\bibfnamefont{S.}~\bibnamefont{Sridhar}} \bibnamefont{and}
  \bibinfo{author}{\bibfnamefont{A.}~\bibnamefont{Kudrolli}},
  \bibinfo{journal}{Phys. Rev. Lett.} \textbf{\bibinfo{volume}{72}},
  \bibinfo{pages}{2175} (\bibinfo{year}{1994}).

\bibitem[{\citenamefont{Hlushchuk et~al.}(2000)\citenamefont{Hlushchuk, Kohler,
  Bauch, Sirko, Bl\"umel, Barth, and St\"ockmann}}]{Hlushchuk2000}
\bibinfo{author}{\bibfnamefont{Y.}~\bibnamefont{Hlushchuk}},
  \bibinfo{author}{\bibfnamefont{A.}~\bibnamefont{Kohler}},
  \bibinfo{author}{\bibfnamefont{S.}~\bibnamefont{Bauch}},
  \bibinfo{author}{\bibfnamefont{L.}~\bibnamefont{Sirko}},
  \bibinfo{author}{\bibfnamefont{R.}~\bibnamefont{Bl\"umel}},
  \bibinfo{author}{\bibfnamefont{M.}~\bibnamefont{Barth}}, \bibnamefont{and}
  \bibinfo{author}{\bibfnamefont{H.-J.} \bibnamefont{St\"ockmann}},
  \bibinfo{journal}{Phys. Rev. E} \textbf{\bibinfo{volume}{61}},
  \bibinfo{pages}{366} (\bibinfo{year}{2000}).

\bibitem[{\citenamefont{Hemmady et~al.}(2005)\citenamefont{Hemmady, Zheng, Ott,
  Antonsen, and Anlage}}]{Hemmady2005}
\bibinfo{author}{\bibfnamefont{S.}~\bibnamefont{Hemmady}},
  \bibinfo{author}{\bibfnamefont{X.}~\bibnamefont{Zheng}},
  \bibinfo{author}{\bibfnamefont{E.}~\bibnamefont{Ott}},
  \bibinfo{author}{\bibfnamefont{T.~M.} \bibnamefont{Antonsen}},
  \bibnamefont{and} \bibinfo{author}{\bibfnamefont{S.~M.}
  \bibnamefont{Anlage}}, \bibinfo{journal}{Phys. Rev. Lett.}
  \textbf{\bibinfo{volume}{94}}, \bibinfo{pages}{014102}
  (\bibinfo{year}{2005}).

\bibitem[{\citenamefont{Hul et~al.}(2005)\citenamefont{Hul, Savytskyy,
  Tymoshchuk, Bauch, and Sirko}}]{Hul2005}
\bibinfo{author}{\bibfnamefont{O.}~\bibnamefont{Hul}},
  \bibinfo{author}{\bibfnamefont{N.}~\bibnamefont{Savytskyy}},
  \bibinfo{author}{\bibfnamefont{O.}~\bibnamefont{Tymoshchuk}},
  \bibinfo{author}{\bibfnamefont{S.}~\bibnamefont{Bauch}}, \bibnamefont{and}
  \bibinfo{author}{\bibfnamefont{L.}~\bibnamefont{Sirko}},
  \bibinfo{journal}{Phys. Rev. E} \textbf{\bibinfo{volume}{72}},
  \bibinfo{pages}{066212} (\bibinfo{year}{2005}).

\bibitem[{\citenamefont{Dietz and Richter}(2015)}]{Dietz2015}
\bibinfo{author}{\bibfnamefont{B.}~\bibnamefont{Dietz}} \bibnamefont{and}
  \bibinfo{author}{\bibfnamefont{A.}~\bibnamefont{Richter}},
  \bibinfo{journal}{Chaos} \textbf{\bibinfo{volume}{25}},
  \bibinfo{pages}{097601} (\bibinfo{year}{2015}).

\bibitem[{\citenamefont{Hul et~al.}(2004)\citenamefont{Hul, Bauch,
  Pako\ifmmode~\acute{n}\else \'{n}\fi{}ski, Savytskyy, \ifmmode~\dot{Z}\else
  \.{Z}\fi{}yczkowski, and Sirko}}]{Hul2004}
\bibinfo{author}{\bibfnamefont{O.}~\bibnamefont{Hul}},
  \bibinfo{author}{\bibfnamefont{S.}~\bibnamefont{Bauch}},
  \bibinfo{author}{\bibfnamefont{P.}~\bibnamefont{Pako\ifmmode~\acute{n}\else
  \'{n}\fi{}ski}}, \bibinfo{author}{\bibfnamefont{N.}~\bibnamefont{Savytskyy}},
  \bibinfo{author}{\bibfnamefont{K.}~\bibnamefont{\ifmmode~\dot{Z}\else
  \.{Z}\fi{}yczkowski}}, \bibnamefont{and}
  \bibinfo{author}{\bibfnamefont{L.}~\bibnamefont{Sirko}},
  \bibinfo{journal}{Phys. Rev. E} \textbf{\bibinfo{volume}{69}},
  \bibinfo{pages}{056205} (\bibinfo{year}{2004}).

\bibitem[{\citenamefont{Hul et~al.}(2012)\citenamefont{Hul, \L{}awniczak,
  Bauch, Sawicki, Ku\ifmmode~\acute{s}\else \'{s}\fi{}, and Sirko}}]{Hul2012}
\bibinfo{author}{\bibfnamefont{O.}~\bibnamefont{Hul}},
  \bibinfo{author}{\bibfnamefont{M.}~\bibnamefont{\L{}awniczak}},
  \bibinfo{author}{\bibfnamefont{S.}~\bibnamefont{Bauch}},
  \bibinfo{author}{\bibfnamefont{A.}~\bibnamefont{Sawicki}},
  \bibinfo{author}{\bibfnamefont{M.}~\bibnamefont{Ku\ifmmode~\acute{s}\else
  \'{s}\fi{}}}, \bibnamefont{and}
  \bibinfo{author}{\bibfnamefont{L.}~\bibnamefont{Sirko}},
  \bibinfo{journal}{Phys. Rev. Lett.} \textbf{\bibinfo{volume}{109}},
  \bibinfo{pages}{040402} (\bibinfo{year}{2012}).

\bibitem[{\citenamefont{Dietz et~al.}(2017{\natexlab{a}})\citenamefont{Dietz,
  Yunko, Bia\l{}ous, Bauch, \L{}awniczak, and Sirko}}]{Bialous2017}
\bibinfo{author}{\bibfnamefont{B.}~\bibnamefont{Dietz}},
  \bibinfo{author}{\bibfnamefont{V.}~\bibnamefont{Yunko}},
  \bibinfo{author}{\bibfnamefont{M.}~\bibnamefont{Bia\l{}ous}},
  \bibinfo{author}{\bibfnamefont{S.}~\bibnamefont{Bauch}},
  \bibinfo{author}{\bibfnamefont{M.}~\bibnamefont{\L{}awniczak}},
  \bibnamefont{and} \bibinfo{author}{\bibfnamefont{L.}~\bibnamefont{Sirko}},
  \bibinfo{journal}{Phys. Rev. E} \textbf{\bibinfo{volume}{95}},
  \bibinfo{pages}{052202} (\bibinfo{year}{2017}{\natexlab{a}}).

\bibitem[{\citenamefont{\L{}awniczak et~al.}(2019)\citenamefont{\L{}awniczak,
  Lipovsk\'y, and Sirko}}]{Lawniczak2019}
\bibinfo{author}{\bibfnamefont{M.}~\bibnamefont{\L{}awniczak}},
  \bibinfo{author}{\bibfnamefont{J.}~\bibnamefont{Lipovsk\'y}},
  \bibnamefont{and} \bibinfo{author}{\bibfnamefont{L.}~\bibnamefont{Sirko}},
  \bibinfo{journal}{Phys. Rev. Lett.} \textbf{\bibinfo{volume}{122}},
  \bibinfo{pages}{140503} (\bibinfo{year}{2019}).

\bibitem[{\citenamefont{Sacha et~al.}(1999)\citenamefont{Sacha, Zakrzewski, and
  Delande}}]{Sacha1999}
\bibinfo{author}{\bibfnamefont{K.}~\bibnamefont{Sacha}},
  \bibinfo{author}{\bibfnamefont{J.}~\bibnamefont{Zakrzewski}},
  \bibnamefont{and} \bibinfo{author}{\bibfnamefont{D.}~\bibnamefont{Delande}},
  \bibinfo{journal}{Phys. Rev. Lett.} \textbf{\bibinfo{volume}{83}},
  \bibinfo{pages}{2922} (\bibinfo{year}{1999}).

\bibitem[{\citenamefont{Ponomarenko et~al.}(2008)\citenamefont{Ponomarenko,
  Schedin, Katsnelson, Yang, Hill, Novoselov, and Geim}}]{Ponomarenko2008}
\bibinfo{author}{\bibfnamefont{L.~A.} \bibnamefont{Ponomarenko}},
  \bibinfo{author}{\bibfnamefont{F.}~\bibnamefont{Schedin}},
  \bibinfo{author}{\bibfnamefont{M.~I.} \bibnamefont{Katsnelson}},
  \bibinfo{author}{\bibfnamefont{R.}~\bibnamefont{Yang}},
  \bibinfo{author}{\bibfnamefont{E.~W.} \bibnamefont{Hill}},
  \bibinfo{author}{\bibfnamefont{K.~S.} \bibnamefont{Novoselov}},
  \bibnamefont{and} \bibinfo{author}{\bibfnamefont{A.~K.} \bibnamefont{Geim}},
  \bibinfo{journal}{Science} \textbf{\bibinfo{volume}{320}},
  \bibinfo{pages}{5874} (\bibinfo{year}{2008}).

\bibitem[{\citenamefont{A\ss~mann et~al.}(2016)\citenamefont{A\ss~mann, Thewes,
  and Fr\"ohlich}}]{Amann2016}
\bibinfo{author}{\bibfnamefont{M.}~\bibnamefont{A\ss~mann}},
  \bibinfo{author}{\bibfnamefont{J.}~\bibnamefont{Thewes}}, \bibnamefont{and}
  \bibinfo{author}{\bibfnamefont{D.}~\bibnamefont{Fr\"ohlich}},
  \bibinfo{journal}{Nat. Mat.} \textbf{\bibinfo{volume}{15}},
  \bibinfo{pages}{741} (\bibinfo{year}{2016}).

\bibitem[{\citenamefont{French et~al.}(1985)\citenamefont{French, Kota, Pandey,
  and Tomsovic}}]{French1985}
\bibinfo{author}{\bibfnamefont{J.~B.} \bibnamefont{French}},
  \bibinfo{author}{\bibfnamefont{V.~K.~B.} \bibnamefont{Kota}},
  \bibinfo{author}{\bibfnamefont{A.}~\bibnamefont{Pandey}}, \bibnamefont{and}
  \bibinfo{author}{\bibfnamefont{S.}~\bibnamefont{Tomsovic}},
  \bibinfo{journal}{Phys. Rev. Lett.} \textbf{\bibinfo{volume}{54}},
  \bibinfo{pages}{2313} (\bibinfo{year}{1985}).

\bibitem[{\citenamefont{Mitchell et~al.}(2010)\citenamefont{Mitchell, Richter,
  and Weidenm\"uller}}]{Mitchell2010}
\bibinfo{author}{\bibfnamefont{G.~E.} \bibnamefont{Mitchell}},
  \bibinfo{author}{\bibfnamefont{A.}~\bibnamefont{Richter}}, \bibnamefont{and}
  \bibinfo{author}{\bibfnamefont{H.~A.} \bibnamefont{Weidenm\"uller}},
  \bibinfo{journal}{Rev. Mod. Phys.} \textbf{\bibinfo{volume}{82}},
  \bibinfo{pages}{2845} (\bibinfo{year}{2010}).

\bibitem[{\citenamefont{So et~al.}(1995)\citenamefont{So, Anlage, Ott, and
  Oerter}}]{So1995}
\bibinfo{author}{\bibfnamefont{P.}~\bibnamefont{So}},
  \bibinfo{author}{\bibfnamefont{S.~M.} \bibnamefont{Anlage}},
  \bibinfo{author}{\bibfnamefont{E.}~\bibnamefont{Ott}}, \bibnamefont{and}
  \bibinfo{author}{\bibfnamefont{R.~N.} \bibnamefont{Oerter}},
  \bibinfo{journal}{Phys. Rev. Lett.} \textbf{\bibinfo{volume}{74}},
  \bibinfo{pages}{2662} (\bibinfo{year}{1995}).

\bibitem[{\citenamefont{Stoffregen et~al.}(1995)\citenamefont{Stoffregen,
  Stein, St\"ockmann, Ku\ifmmode~\acute{s}\else \'{s}\fi{}, and
  Haake}}]{Stoffregen1995}
\bibinfo{author}{\bibfnamefont{U.}~\bibnamefont{Stoffregen}},
  \bibinfo{author}{\bibfnamefont{J.}~\bibnamefont{Stein}},
  \bibinfo{author}{\bibfnamefont{H.-J.} \bibnamefont{St\"ockmann}},
  \bibinfo{author}{\bibfnamefont{M.}~\bibnamefont{Ku\ifmmode~\acute{s}\else
  \'{s}\fi{}}}, \bibnamefont{and}
  \bibinfo{author}{\bibfnamefont{F.}~\bibnamefont{Haake}},
  \bibinfo{journal}{Phys. Rev. Lett.} \textbf{\bibinfo{volume}{74}},
  \bibinfo{pages}{2666} (\bibinfo{year}{1995}).

\bibitem[{\citenamefont{Dietz et~al.}(2007)\citenamefont{Dietz, Friedrich,
  Harney, Miski-Oglu, Richter, Sch\"{a}fer, and Weidenm\"{u}ller}}]{Dietz2007a}
\bibinfo{author}{\bibfnamefont{B.}~\bibnamefont{Dietz}},
  \bibinfo{author}{\bibfnamefont{T.}~\bibnamefont{Friedrich}},
  \bibinfo{author}{\bibfnamefont{H.~L.} \bibnamefont{Harney}},
  \bibinfo{author}{\bibfnamefont{M.}~\bibnamefont{Miski-Oglu}},
  \bibinfo{author}{\bibfnamefont{A.}~\bibnamefont{Richter}},
  \bibinfo{author}{\bibfnamefont{F.}~\bibnamefont{Sch\"{a}fer}},
  \bibnamefont{and} \bibinfo{author}{\bibfnamefont{H.~A.}
  \bibnamefont{Weidenm\"{u}ller}}, \bibinfo{journal}{Phys. Rev. Lett.}
  \textbf{\bibinfo{volume}{98}}, \bibinfo{eid}{074103} (\bibinfo{year}{2007}).

\bibitem[{\citenamefont{\L{}awniczak et~al.}(2010)\citenamefont{\L{}awniczak,
  Bauch, Hul, and Sirko}}]{Lawniczak2010}
\bibinfo{author}{\bibfnamefont{M.}~\bibnamefont{\L{}awniczak}},
  \bibinfo{author}{\bibfnamefont{S.}~\bibnamefont{Bauch}},
  \bibinfo{author}{\bibfnamefont{O.}~\bibnamefont{Hul}}, \bibnamefont{and}
  \bibinfo{author}{\bibfnamefont{L.}~\bibnamefont{Sirko}},
  \bibinfo{journal}{Phys. Rev. E} \textbf{\bibinfo{volume}{81}},
  \bibinfo{pages}{046204} (\bibinfo{year}{2010}).

\bibitem[{\citenamefont{Bia\l{}ous
  et~al.}(2016{\natexlab{a}})\citenamefont{Bia\l{}ous, Yunko, Bauch,
  \L{}awniczak, Dietz, and Sirko}}]{Bialous2016}
\bibinfo{author}{\bibfnamefont{M.}~\bibnamefont{Bia\l{}ous}},
  \bibinfo{author}{\bibfnamefont{V.}~\bibnamefont{Yunko}},
  \bibinfo{author}{\bibfnamefont{S.}~\bibnamefont{Bauch}},
  \bibinfo{author}{\bibfnamefont{M.}~\bibnamefont{\L{}awniczak}},
  \bibinfo{author}{\bibfnamefont{B.}~\bibnamefont{Dietz}}, \bibnamefont{and}
  \bibinfo{author}{\bibfnamefont{L.}~\bibnamefont{Sirko}},
  \bibinfo{journal}{Phys. Rev. Lett.} \textbf{\bibinfo{volume}{117}},
  \bibinfo{pages}{144101} (\bibinfo{year}{2016}{\natexlab{a}}).

\bibitem[{\citenamefont{Rehemanjiang et~al.}(2018)\citenamefont{Rehemanjiang,
  Richter, Kuhl, and St\"ockmann}}]{Rehemanjiang2018}
\bibinfo{author}{\bibfnamefont{A.}~\bibnamefont{Rehemanjiang}},
  \bibinfo{author}{\bibfnamefont{M.}~\bibnamefont{Richter}},
  \bibinfo{author}{\bibfnamefont{U.}~\bibnamefont{Kuhl}}, \bibnamefont{and}
  \bibinfo{author}{\bibfnamefont{H.-J.} \bibnamefont{St\"ockmann}},
  \bibinfo{journal}{Phys. Rev. E} \textbf{\bibinfo{volume}{97}},
  \bibinfo{pages}{022204} (\bibinfo{year}{2018}).

\bibitem[{\citenamefont{L{}awniczak and Sirko}(2019)}]{Lawniczak2019b}
\bibinfo{author}{\bibfnamefont{M.}~\bibnamefont{L{}awniczak}} \bibnamefont{and}
  \bibinfo{author}{\bibfnamefont{L.}~\bibnamefont{Sirko}},
  \bibinfo{journal}{Sci. Rep.} \textbf{\bibinfo{volume}{9}},
  \bibinfo{pages}{5630} (\bibinfo{year}{2019}).

\bibitem[{\citenamefont{Lu et~al.}(2020)\citenamefont{Lu, Che, Zhang, and
  Dietz}}]{Lu2020}
\bibinfo{author}{\bibfnamefont{J.}~\bibnamefont{Lu}},
  \bibinfo{author}{\bibfnamefont{J.}~\bibnamefont{Che}},
  \bibinfo{author}{\bibfnamefont{X.}~\bibnamefont{Zhang}}, \bibnamefont{and}
  \bibinfo{author}{\bibfnamefont{B.}~\bibnamefont{Dietz}},
  \bibinfo{journal}{Phys. Rev. E} \textbf{\bibinfo{volume}{102}},
  \bibinfo{pages}{022309} (\bibinfo{year}{2020}).

\bibitem[{\citenamefont{Yunko et~al.}(2020)\citenamefont{Yunko, Bia\l{}ous, and
  Sirko}}]{Yunko2020}
\bibinfo{author}{\bibfnamefont{V.}~\bibnamefont{Yunko}},
  \bibinfo{author}{\bibfnamefont{M.}~\bibnamefont{Bia\l{}ous}},
  \bibnamefont{and} \bibinfo{author}{\bibfnamefont{L.}~\bibnamefont{Sirko}},
  \bibinfo{journal}{Phys. Rev. E} \textbf{\bibinfo{volume}{102}},
  \bibinfo{pages}{012210} (\bibinfo{year}{2020}).

\bibitem[{\citenamefont{\L{}awniczak et~al.}(2020)\citenamefont{\L{}awniczak,
  van Tiggelen, and Sirko}}]{Lawniczak2020}
\bibinfo{author}{\bibfnamefont{M.}~\bibnamefont{\L{}awniczak}},
  \bibinfo{author}{\bibfnamefont{B.}~\bibnamefont{van Tiggelen}},
  \bibnamefont{and} \bibinfo{author}{\bibfnamefont{L.}~\bibnamefont{Sirko}},
  \bibinfo{journal}{Phys. Rev. E} \textbf{\bibinfo{volume}{102}},
  \bibinfo{pages}{052214} (\bibinfo{year}{2020}).

\bibitem[{\citenamefont{Pandey and Shukla}(1991)}]{Pandey1991}
\bibinfo{author}{\bibfnamefont{A.}~\bibnamefont{Pandey}} \bibnamefont{and}
  \bibinfo{author}{\bibfnamefont{P.}~\bibnamefont{Shukla}},
  \bibinfo{journal}{J. Phys. A} \textbf{\bibinfo{volume}{24}},
  \bibinfo{pages}{3907} (\bibinfo{year}{1991}).

\bibitem[{\citenamefont{Dietz}(1991)}]{Dietz1991}
\bibinfo{author}{\bibfnamefont{B.}~\bibnamefont{Dietz}}, Ph.D. thesis,
  \bibinfo{school}{Fachbereich Physik der Universit\"at-Gesamthochschule Essen}
  (\bibinfo{year}{1991}).

\bibitem[{\citenamefont{Lenz and \ifmmode~\dot{Z}\else
  \.{Z}\fi{}yczkowski}(1992)}]{Lenz1992}
\bibinfo{author}{\bibfnamefont{G.}~\bibnamefont{Lenz}} \bibnamefont{and}
  \bibinfo{author}{\bibfnamefont{K.}~\bibnamefont{\ifmmode~\dot{Z}\else
  \.{Z}\fi{}yczkowski}}, \bibinfo{journal}{J. Phys. A}
  \textbf{\bibinfo{volume}{25}}, \bibinfo{pages}{5539} (\bibinfo{year}{1992}).

\bibitem[{\citenamefont{Schierenberg et~al.}(2012)\citenamefont{Schierenberg,
  Bruckmann, and Wettig}}]{Schierenberg2012}
\bibinfo{author}{\bibfnamefont{S.}~\bibnamefont{Schierenberg}},
  \bibinfo{author}{\bibfnamefont{F.}~\bibnamefont{Bruckmann}},
  \bibnamefont{and} \bibinfo{author}{\bibfnamefont{T.}~\bibnamefont{Wettig}},
  \bibinfo{journal}{Phys. Rev. E} \textbf{\bibinfo{volume}{85}},
  \bibinfo{pages}{061130} (\bibinfo{year}{2012}).

\bibitem[{\citenamefont{Dietz et~al.}(2009)\citenamefont{Dietz, Friedrich,
  Harney, Miski-Oglu, Richter, Sch\"afer, Verbaarschot, and
  Weidenm\"uller}}]{Dietz2009a}
\bibinfo{author}{\bibfnamefont{B.}~\bibnamefont{Dietz}},
  \bibinfo{author}{\bibfnamefont{T.}~\bibnamefont{Friedrich}},
  \bibinfo{author}{\bibfnamefont{H.~L.} \bibnamefont{Harney}},
  \bibinfo{author}{\bibfnamefont{M.}~\bibnamefont{Miski-Oglu}},
  \bibinfo{author}{\bibfnamefont{A.}~\bibnamefont{Richter}},
  \bibinfo{author}{\bibfnamefont{F.}~\bibnamefont{Sch\"afer}},
  \bibinfo{author}{\bibfnamefont{J.}~\bibnamefont{Verbaarschot}},
  \bibnamefont{and} \bibinfo{author}{\bibfnamefont{H.~A.}
  \bibnamefont{Weidenm\"uller}}, \bibinfo{journal}{Phys. Rev. Lett.}
  \textbf{\bibinfo{volume}{103}}, \bibinfo{pages}{064101}
  (\bibinfo{year}{2009}).

\bibitem[{\citenamefont{Dietz et~al.}(2019)\citenamefont{Dietz, Klaus,
  Miski-Oglu, Richter, and Wunderle}}]{Dietz2019}
\bibinfo{author}{\bibfnamefont{B.}~\bibnamefont{Dietz}},
  \bibinfo{author}{\bibfnamefont{T.}~\bibnamefont{Klaus}},
  \bibinfo{author}{\bibfnamefont{M.}~\bibnamefont{Miski-Oglu}},
  \bibinfo{author}{\bibfnamefont{A.}~\bibnamefont{Richter}}, \bibnamefont{and}
  \bibinfo{author}{\bibfnamefont{M.}~\bibnamefont{Wunderle}},
  \bibinfo{journal}{Phys. Rev. Lett.} \textbf{\bibinfo{volume}{123}},
  \bibinfo{pages}{174101} (\bibinfo{year}{2019}).

\bibitem[{\citenamefont{Bia\l{}ous et~al.}(2020)\citenamefont{Bia\l{}ous,
  Dietz, and Sirko}}]{Bialous2020}
\bibinfo{author}{\bibfnamefont{M.}~\bibnamefont{Bia\l{}ous}},
  \bibinfo{author}{\bibfnamefont{B.}~\bibnamefont{Dietz}}, \bibnamefont{and}
  \bibinfo{author}{\bibfnamefont{L.}~\bibnamefont{Sirko}},
  \bibinfo{journal}{Phys. Rev. E} \textbf{\bibinfo{volume}{102}},
  \bibinfo{pages}{042206} (\bibinfo{year}{2020}).

\bibitem[{\citenamefont{Dietz et~al.}(2011)\citenamefont{Dietz, Harney,
  Kirillov, Miski-Oglu, Richter, and Sch\"afer}}]{Dietz2011}
\bibinfo{author}{\bibfnamefont{B.}~\bibnamefont{Dietz}},
  \bibinfo{author}{\bibfnamefont{H.~L.} \bibnamefont{Harney}},
  \bibinfo{author}{\bibfnamefont{O.~N.} \bibnamefont{Kirillov}},
  \bibinfo{author}{\bibfnamefont{M.}~\bibnamefont{Miski-Oglu}},
  \bibinfo{author}{\bibfnamefont{A.}~\bibnamefont{Richter}}, \bibnamefont{and}
  \bibinfo{author}{\bibfnamefont{F.}~\bibnamefont{Sch\"afer}},
  \bibinfo{journal}{Phys. Rev. Lett.} \textbf{\bibinfo{volume}{106}},
  \bibinfo{pages}{150403} (\bibinfo{year}{2011}).

\bibitem[{\citenamefont{Bittner et~al.}(2012)\citenamefont{Bittner, Dietz,
  G\"unther, Harney, Miski-Oglu, Richter, and Sch\"afer}}]{Dietz2012}
\bibinfo{author}{\bibfnamefont{S.}~\bibnamefont{Bittner}},
  \bibinfo{author}{\bibfnamefont{B.}~\bibnamefont{Dietz}},
  \bibinfo{author}{\bibfnamefont{U.}~\bibnamefont{G\"unther}},
  \bibinfo{author}{\bibfnamefont{H.~L.} \bibnamefont{Harney}},
  \bibinfo{author}{\bibfnamefont{M.}~\bibnamefont{Miski-Oglu}},
  \bibinfo{author}{\bibfnamefont{A.}~\bibnamefont{Richter}}, \bibnamefont{and}
  \bibinfo{author}{\bibfnamefont{F.}~\bibnamefont{Sch\"afer}},
  \bibinfo{journal}{Phys. Rev. Lett.} \textbf{\bibinfo{volume}{108}},
  \bibinfo{pages}{024101} (\bibinfo{year}{2012}).

\bibitem[{\citenamefont{Dietz et~al.}(2014)\citenamefont{Dietz, Guhr, Gutkin,
  Miski-Oglu, and Richter}}]{Dietz2014}
\bibinfo{author}{\bibfnamefont{B.}~\bibnamefont{Dietz}},
  \bibinfo{author}{\bibfnamefont{T.}~\bibnamefont{Guhr}},
  \bibinfo{author}{\bibfnamefont{B.}~\bibnamefont{Gutkin}},
  \bibinfo{author}{\bibfnamefont{M.}~\bibnamefont{Miski-Oglu}},
  \bibnamefont{and} \bibinfo{author}{\bibfnamefont{A.}~\bibnamefont{Richter}},
  \bibinfo{journal}{Phys. Rev. E} \textbf{\bibinfo{volume}{90}},
  \bibinfo{pages}{022903} (\bibinfo{year}{2014}).

\bibitem[{\citenamefont{Dietz et~al.}(2016)\citenamefont{Dietz, Klaus,
  Miski-Oglu, Richter, Wunderle, and Bouazza}}]{Dietz2016}
\bibinfo{author}{\bibfnamefont{B.}~\bibnamefont{Dietz}},
  \bibinfo{author}{\bibfnamefont{T.}~\bibnamefont{Klaus}},
  \bibinfo{author}{\bibfnamefont{M.}~\bibnamefont{Miski-Oglu}},
  \bibinfo{author}{\bibfnamefont{A.}~\bibnamefont{Richter}},
  \bibinfo{author}{\bibfnamefont{M.}~\bibnamefont{Wunderle}}, \bibnamefont{and}
  \bibinfo{author}{\bibfnamefont{C.}~\bibnamefont{Bouazza}},
  \bibinfo{journal}{Phys. Rev. Lett.} \textbf{\bibinfo{volume}{116}},
  \bibinfo{pages}{023901} (\bibinfo{year}{2016}).

\bibitem[{\citenamefont{Bohigas and Pato}(2004)}]{Bohigas2004}
\bibinfo{author}{\bibfnamefont{O.}~\bibnamefont{Bohigas}} \bibnamefont{and}
  \bibinfo{author}{\bibfnamefont{M.~P.} \bibnamefont{Pato}},
  \bibinfo{journal}{Phys. Lett. B} \textbf{\bibinfo{volume}{595}},
  \bibinfo{pages}{171} (\bibinfo{year}{2004}).

\bibitem[{\citenamefont{Agvaanluvsan
  et~al.}(2003{\natexlab{a}})\citenamefont{Agvaanluvsan, Mitchell,
  Schriner~Jr., and Pato}}]{Agvaanluvsan2003}
\bibinfo{author}{\bibfnamefont{U.}~\bibnamefont{Agvaanluvsan}},
  \bibinfo{author}{\bibfnamefont{G.~E.} \bibnamefont{Mitchell}},
  \bibinfo{author}{\bibfnamefont{J.~F.} \bibnamefont{Schriner~Jr.}},
  \bibnamefont{and} \bibinfo{author}{\bibfnamefont{M.~P.} \bibnamefont{Pato}},
  \bibinfo{journal}{Nuclear Instruments and Methods in Physics Research A}
  \textbf{\bibinfo{volume}{498}}, \bibinfo{pages}{459}
  (\bibinfo{year}{2003}{\natexlab{a}}).

\bibitem[{\citenamefont{Agvaanluvsan
  et~al.}(2003{\natexlab{b}})\citenamefont{Agvaanluvsan, Mitchell, Shriner, and
  Pato}}]{Agvaanluvsan2003a}
\bibinfo{author}{\bibfnamefont{U.}~\bibnamefont{Agvaanluvsan}},
  \bibinfo{author}{\bibfnamefont{G.~E.} \bibnamefont{Mitchell}},
  \bibinfo{author}{\bibfnamefont{J.~F.} \bibnamefont{Shriner}},
  \bibnamefont{and} \bibinfo{author}{\bibfnamefont{M.}~\bibnamefont{Pato}},
  \bibinfo{journal}{Phys. Rev. C} \textbf{\bibinfo{volume}{67}},
  \bibinfo{pages}{064608} (\bibinfo{year}{2003}{\natexlab{b}}).

\bibitem[{\citenamefont{Bohigas and Pato}(2006)}]{Bohigas2006}
\bibinfo{author}{\bibfnamefont{O.}~\bibnamefont{Bohigas}} \bibnamefont{and}
  \bibinfo{author}{\bibfnamefont{M.~P.} \bibnamefont{Pato}},
  \bibinfo{journal}{Phys. Rev. E} \textbf{\bibinfo{volume}{74}},
  \bibinfo{pages}{036212} (\bibinfo{year}{2006}).

\bibitem[{\citenamefont{Liou et~al.}(1972)\citenamefont{Liou, Camarda, and
  Rahn}}]{Liou1972}
\bibinfo{author}{\bibfnamefont{H.~I.} \bibnamefont{Liou}},
  \bibinfo{author}{\bibfnamefont{H.~S.} \bibnamefont{Camarda}},
  \bibnamefont{and} \bibinfo{author}{\bibfnamefont{F.}~\bibnamefont{Rahn}},
  \bibinfo{journal}{Phys. Rev. C} \textbf{\bibinfo{volume}{5}},
  \bibinfo{pages}{1002} (\bibinfo{year}{1972}).

\bibitem[{\citenamefont{Brody et~al.}(1981)\citenamefont{Brody, Flores, French,
  Mello, Pandey, and Wong}}]{Brody1981}
\bibinfo{author}{\bibfnamefont{T.~A.} \bibnamefont{Brody}},
  \bibinfo{author}{\bibfnamefont{J.}~\bibnamefont{Flores}},
  \bibinfo{author}{\bibfnamefont{J.~B.} \bibnamefont{French}},
  \bibinfo{author}{\bibfnamefont{P.~A.} \bibnamefont{Mello}},
  \bibinfo{author}{\bibfnamefont{A.}~\bibnamefont{Pandey}}, \bibnamefont{and}
  \bibinfo{author}{\bibfnamefont{S.~S.~M.} \bibnamefont{Wong}},
  \bibinfo{journal}{Rev. Mod. Phys.} \textbf{\bibinfo{volume}{53}},
  \bibinfo{pages}{385} (\bibinfo{year}{1981}).

\bibitem[{\citenamefont{Enders et~al.}(2000)\citenamefont{Enders, Guhr, Huxel,
  {von Neumann-Cosel}, Rangacharyulu, and Richter}}]{Enders2000}
\bibinfo{author}{\bibfnamefont{J.}~\bibnamefont{Enders}},
  \bibinfo{author}{\bibfnamefont{T.}~\bibnamefont{Guhr}},
  \bibinfo{author}{\bibfnamefont{N.}~\bibnamefont{Huxel}},
  \bibinfo{author}{\bibfnamefont{P.}~\bibnamefont{{von Neumann-Cosel}}},
  \bibinfo{author}{\bibfnamefont{C.}~\bibnamefont{Rangacharyulu}},
  \bibnamefont{and} \bibinfo{author}{\bibfnamefont{A.}~\bibnamefont{Richter}},
  \bibinfo{journal}{Physics Letters B} \textbf{\bibinfo{volume}{486}},
  \bibinfo{pages}{273 } (\bibinfo{year}{2000}).

\bibitem[{\citenamefont{Enders et~al.}(2004)\citenamefont{Enders, Guhr, Heine,
  von Neumann Cosel, Ponomarev, Richter, and Wambach}}]{Enders2004}
\bibinfo{author}{\bibfnamefont{J.}~\bibnamefont{Enders}},
  \bibinfo{author}{\bibfnamefont{T.}~\bibnamefont{Guhr}},
  \bibinfo{author}{\bibfnamefont{A.}~\bibnamefont{Heine}},
  \bibinfo{author}{\bibfnamefont{P.}~\bibnamefont{von Neumann Cosel}},
  \bibinfo{author}{\bibfnamefont{V.}~\bibnamefont{Ponomarev}},
  \bibinfo{author}{\bibfnamefont{A.}~\bibnamefont{Richter}}, \bibnamefont{and}
  \bibinfo{author}{\bibfnamefont{J.}~\bibnamefont{Wambach}},
  \bibinfo{journal}{Nuclear Physics A} \textbf{\bibinfo{volume}{741}},
  \bibinfo{pages}{3 } (\bibinfo{year}{2004}).

\bibitem[{\citenamefont{Molina et~al.}(2007)\citenamefont{Molina, Retamosa,
  Muñoz, Relaño, and Faleiro}}]{Molina2007}
\bibinfo{author}{\bibfnamefont{R.}~\bibnamefont{Molina}},
  \bibinfo{author}{\bibfnamefont{J.}~\bibnamefont{Retamosa}},
  \bibinfo{author}{\bibfnamefont{L.}~\bibnamefont{Muñoz}},
  \bibinfo{author}{\bibfnamefont{A.}~\bibnamefont{Relaño}}, \bibnamefont{and}
  \bibinfo{author}{\bibfnamefont{E.}~\bibnamefont{Faleiro}},
  \bibinfo{journal}{Physics Letters B} \textbf{\bibinfo{volume}{644}},
  \bibinfo{pages}{25 } (\bibinfo{year}{2007}).

\bibitem[{\citenamefont{Frisch et~al.}(2014)\citenamefont{Frisch, Mark, Aikawa,
  Ferlaino, Bohn, Makrides, Petrov, and Kotochigova}}]{Frisch2014}
\bibinfo{author}{\bibfnamefont{A.}~\bibnamefont{Frisch}},
  \bibinfo{author}{\bibfnamefont{M.}~\bibnamefont{Mark}},
  \bibinfo{author}{\bibfnamefont{K.}~\bibnamefont{Aikawa}},
  \bibinfo{author}{\bibfnamefont{F.}~\bibnamefont{Ferlaino}},
  \bibinfo{author}{\bibfnamefont{J.~L.} \bibnamefont{Bohn}},
  \bibinfo{author}{\bibfnamefont{C.}~\bibnamefont{Makrides}},
  \bibinfo{author}{\bibfnamefont{A.}~\bibnamefont{Petrov}}, \bibnamefont{and}
  \bibinfo{author}{\bibfnamefont{S.}~\bibnamefont{Kotochigova}},
  \bibinfo{journal}{Nature} \textbf{\bibinfo{volume}{507}},
  \bibinfo{pages}{474} (\bibinfo{year}{2014}).

\bibitem[{\citenamefont{Mur-Petit and Molina}(2015)}]{Mur2015}
\bibinfo{author}{\bibfnamefont{J.}~\bibnamefont{Mur-Petit}} \bibnamefont{and}
  \bibinfo{author}{\bibfnamefont{R.~A.} \bibnamefont{Molina}},
  \bibinfo{journal}{Phys. Rev. E} \textbf{\bibinfo{volume}{92}},
  \bibinfo{pages}{042906} (\bibinfo{year}{2015}).

\bibitem[{\citenamefont{Bia\l{}ous
  et~al.}(2016{\natexlab{b}})\citenamefont{Bia\l{}ous, Yunko, Bauch,
  \L{}awniczak, Dietz, and Sirko}}]{Bialous2016a}
\bibinfo{author}{\bibfnamefont{M.}~\bibnamefont{Bia\l{}ous}},
  \bibinfo{author}{\bibfnamefont{V.}~\bibnamefont{Yunko}},
  \bibinfo{author}{\bibfnamefont{S.}~\bibnamefont{Bauch}},
  \bibinfo{author}{\bibfnamefont{M.}~\bibnamefont{\L{}awniczak}},
  \bibinfo{author}{\bibfnamefont{B.}~\bibnamefont{Dietz}}, \bibnamefont{and}
  \bibinfo{author}{\bibfnamefont{L.}~\bibnamefont{Sirko}},
  \bibinfo{journal}{Phys. Rev. E} \textbf{\bibinfo{volume}{94}},
  \bibinfo{pages}{042211} (\bibinfo{year}{2016}{\natexlab{b}}).

\bibitem[{\citenamefont{\L{}awniczak et~al.}(2018)\citenamefont{\L{}awniczak,
  Bia\l{}ous, Yunko, Bauch, and Sirko}}]{Lawniczak2018}
\bibinfo{author}{\bibfnamefont{M.}~\bibnamefont{\L{}awniczak}},
  \bibinfo{author}{\bibfnamefont{M.}~\bibnamefont{Bia\l{}ous}},
  \bibinfo{author}{\bibfnamefont{V.}~\bibnamefont{Yunko}},
  \bibinfo{author}{\bibfnamefont{S.}~\bibnamefont{Bauch}}, \bibnamefont{and}
  \bibinfo{author}{\bibfnamefont{L.}~\bibnamefont{Sirko}},
  \bibinfo{journal}{Phys. Rev. E} \textbf{\bibinfo{volume}{98}},
  \bibinfo{pages}{012206} (\bibinfo{year}{2018}).

\bibitem[{\citenamefont{Bia\l{}ous et~al.}(2019)\citenamefont{Bia\l{}ous,
  Dietz, and Sirko}}]{Bialous2019}
\bibinfo{author}{\bibfnamefont{M.}~\bibnamefont{Bia\l{}ous}},
  \bibinfo{author}{\bibfnamefont{B.}~\bibnamefont{Dietz}}, \bibnamefont{and}
  \bibinfo{author}{\bibfnamefont{L.}~\bibnamefont{Sirko}},
  \bibinfo{journal}{Phys. Rev. E} \textbf{\bibinfo{volume}{100}},
  \bibinfo{pages}{012210} (\bibinfo{year}{2019}).

\bibitem[{\citenamefont{St{\"o}ckmann}(2000)}]{StoeckmannBuch2000}
\bibinfo{author}{\bibfnamefont{H.-J.} \bibnamefont{St{\"o}ckmann}},
  \emph{\bibinfo{title}{Quantum Chaos: An Introduction}}
  (\bibinfo{publisher}{Cambridge University Press},
  \bibinfo{address}{Cambridge}, \bibinfo{year}{2000}).

\bibitem[{\citenamefont{Haake}(2001)}]{Haake2001}
\bibinfo{author}{\bibfnamefont{F.}~\bibnamefont{Haake}},
  \emph{\bibinfo{title}{Quantum Signatures of Chaos}}
  (\bibinfo{publisher}{Springer-Verlag}, \bibinfo{address}{Heidelberg},
  \bibinfo{year}{2001}).

\bibitem[{\citenamefont{Rela\~no et~al.}(2002)\citenamefont{Rela\~no, G\'omez,
  Molina, Retamosa, and Faleiro}}]{Relano2002}
\bibinfo{author}{\bibfnamefont{A.}~\bibnamefont{Rela\~no}},
  \bibinfo{author}{\bibfnamefont{J.~M.~G.} \bibnamefont{G\'omez}},
  \bibinfo{author}{\bibfnamefont{R.~A.} \bibnamefont{Molina}},
  \bibinfo{author}{\bibfnamefont{J.}~\bibnamefont{Retamosa}}, \bibnamefont{and}
  \bibinfo{author}{\bibfnamefont{E.}~\bibnamefont{Faleiro}},
  \bibinfo{journal}{Phys. Rev. Lett.} \textbf{\bibinfo{volume}{89}},
  \bibinfo{pages}{244102} (\bibinfo{year}{2002}).

\bibitem[{\citenamefont{Faleiro et~al.}(2004)\citenamefont{Faleiro, G\'omez,
  Molina, Mu\~noz, Rela\~no, and Retamosa}}]{Faleiro2004}
\bibinfo{author}{\bibfnamefont{E.}~\bibnamefont{Faleiro}},
  \bibinfo{author}{\bibfnamefont{J.~M.~G.} \bibnamefont{G\'omez}},
  \bibinfo{author}{\bibfnamefont{R.~A.} \bibnamefont{Molina}},
  \bibinfo{author}{\bibfnamefont{L.}~\bibnamefont{Mu\~noz}},
  \bibinfo{author}{\bibfnamefont{A.}~\bibnamefont{Rela\~no}}, \bibnamefont{and}
  \bibinfo{author}{\bibfnamefont{J.}~\bibnamefont{Retamosa}},
  \bibinfo{journal}{Phys. Rev. Lett.} \textbf{\bibinfo{volume}{93}},
  \bibinfo{pages}{244101} (\bibinfo{year}{2004}).

\bibitem[{\citenamefont{Riser et~al.}(2017)\citenamefont{Riser, Osipov, and
  Kanzieper}}]{Riser2017}
\bibinfo{author}{\bibfnamefont{R.}~\bibnamefont{Riser}},
  \bibinfo{author}{\bibfnamefont{V.~A.} \bibnamefont{Osipov}},
  \bibnamefont{and}
  \bibinfo{author}{\bibfnamefont{E.}~\bibnamefont{Kanzieper}},
  \bibinfo{journal}{Phys. Rev. Lett.} \textbf{\bibinfo{volume}{118}},
  \bibinfo{pages}{204101} (\bibinfo{year}{2017}).

\bibitem[{\citenamefont{Dietz et~al.}(2017{\natexlab{b}})\citenamefont{Dietz,
  Yunko, Bia\l{}ous, Bauch, \L{}awniczak, and Sirko}}]{Dietz2017}
\bibinfo{author}{\bibfnamefont{B.}~\bibnamefont{Dietz}},
  \bibinfo{author}{\bibfnamefont{V.}~\bibnamefont{Yunko}},
  \bibinfo{author}{\bibfnamefont{M.}~\bibnamefont{Bia\l{}ous}},
  \bibinfo{author}{\bibfnamefont{S.}~\bibnamefont{Bauch}},
  \bibinfo{author}{\bibfnamefont{M.}~\bibnamefont{\L{}awniczak}},
  \bibnamefont{and} \bibinfo{author}{\bibfnamefont{L.}~\bibnamefont{Sirko}},
  \bibinfo{journal}{Phys. Rev. E} \textbf{\bibinfo{volume}{95}},
  \bibinfo{pages}{052202} (\bibinfo{year}{2017}{\natexlab{b}}).

\bibitem[{\citenamefont{G\'omez et~al.}(2005)\citenamefont{G\'omez, Rela\~no,
  Retamosa, Faleiro, Salasnich, Vrani\ifmmode~\check{c}\else \v{c}\fi{}ar, and
  Robnik}}]{Gomez2005}
\bibinfo{author}{\bibfnamefont{J.~M.~G.} \bibnamefont{G\'omez}},
  \bibinfo{author}{\bibfnamefont{A.}~\bibnamefont{Rela\~no}},
  \bibinfo{author}{\bibfnamefont{J.}~\bibnamefont{Retamosa}},
  \bibinfo{author}{\bibfnamefont{E.}~\bibnamefont{Faleiro}},
  \bibinfo{author}{\bibfnamefont{L.}~\bibnamefont{Salasnich}},
  \bibinfo{author}{\bibfnamefont{M.}~\bibnamefont{Vrani\ifmmode~\check{c}\else
  \v{c}\fi{}ar}}, \bibnamefont{and}
  \bibinfo{author}{\bibfnamefont{M.}~\bibnamefont{Robnik}},
  \bibinfo{journal}{Phys. Rev. Lett.} \textbf{\bibinfo{volume}{94}},
  \bibinfo{pages}{084101} (\bibinfo{year}{2005}).

\bibitem[{\citenamefont{Salasnich}(2005)}]{Salasnich2005}
\bibinfo{author}{\bibfnamefont{L.}~\bibnamefont{Salasnich}},
  \bibinfo{journal}{Phys. Rev. E} \textbf{\bibinfo{volume}{71}},
  \bibinfo{pages}{047202} (\bibinfo{year}{2005}).

\bibitem[{\citenamefont{Santhanam and Bandyopadhyay}(2005)}]{Santhanam2005}
\bibinfo{author}{\bibfnamefont{M.~S.} \bibnamefont{Santhanam}}
  \bibnamefont{and} \bibinfo{author}{\bibfnamefont{J.~N.}
  \bibnamefont{Bandyopadhyay}}, \bibinfo{journal}{Phys. Rev. Lett.}
  \textbf{\bibinfo{volume}{95}}, \bibinfo{pages}{114101}
  (\bibinfo{year}{2005}).

\bibitem[{\citenamefont{Rela\~no}(2008)}]{Relano2008}
\bibinfo{author}{\bibfnamefont{A.}~\bibnamefont{Rela\~no}},
  \bibinfo{journal}{Phys. Rev. Lett.} \textbf{\bibinfo{volume}{100}},
  \bibinfo{pages}{224101} (\bibinfo{year}{2008}).

\bibitem[{\citenamefont{Faleiro et~al.}(2006)\citenamefont{Faleiro, Kuhl,
  Molina, Muñoz, Relaño, and Retamosa}}]{Faleiro2006}
\bibinfo{author}{\bibfnamefont{E.}~\bibnamefont{Faleiro}},
  \bibinfo{author}{\bibfnamefont{U.}~\bibnamefont{Kuhl}},
  \bibinfo{author}{\bibfnamefont{R.}~\bibnamefont{Molina}},
  \bibinfo{author}{\bibfnamefont{L.}~\bibnamefont{Muñoz}},
  \bibinfo{author}{\bibfnamefont{A.}~\bibnamefont{Relaño}}, \bibnamefont{and}
  \bibinfo{author}{\bibfnamefont{J.}~\bibnamefont{Retamosa}},
  \bibinfo{journal}{Physics Letters A} \textbf{\bibinfo{volume}{358}},
  \bibinfo{pages}{251 } (\bibinfo{year}{2006}).

\bibitem[{\citenamefont{Dietz et~al.}(2010)\citenamefont{Dietz, Friedrich,
  Harney, Miski-Oglu, Richter, Sch\"afer, and Weidenm\"uller}}]{Dietz2010}
\bibinfo{author}{\bibfnamefont{B.}~\bibnamefont{Dietz}},
  \bibinfo{author}{\bibfnamefont{T.}~\bibnamefont{Friedrich}},
  \bibinfo{author}{\bibfnamefont{H.~L.} \bibnamefont{Harney}},
  \bibinfo{author}{\bibfnamefont{M.}~\bibnamefont{Miski-Oglu}},
  \bibinfo{author}{\bibfnamefont{A.}~\bibnamefont{Richter}},
  \bibinfo{author}{\bibfnamefont{F.}~\bibnamefont{Sch\"afer}},
  \bibnamefont{and} \bibinfo{author}{\bibfnamefont{H.~A.}
  \bibnamefont{Weidenm\"uller}}, \bibinfo{journal}{Phys. Rev. E}
  \textbf{\bibinfo{volume}{81}}, \bibinfo{pages}{036205}
  (\bibinfo{year}{2010}).

\bibitem[{\citenamefont{Bohigas et~al.}(1995)\citenamefont{Bohigas, Giannoni,
  de~Almeidaz, and Schmit}}]{Bohigas1995}
\bibinfo{author}{\bibfnamefont{O.}~\bibnamefont{Bohigas}},
  \bibinfo{author}{\bibfnamefont{M.-J.} \bibnamefont{Giannoni}},
  \bibinfo{author}{\bibfnamefont{A.~M.~O.} \bibnamefont{de~Almeidaz}},
  \bibnamefont{and} \bibinfo{author}{\bibfnamefont{C.}~\bibnamefont{Schmit}},
  \bibinfo{journal}{Nonlinearity} \textbf{\bibinfo{volume}{8}},
  \bibinfo{pages}{203} (\bibinfo{year}{1995}).

\bibitem[{\citenamefont{Bohigas et~al.}(1999)\citenamefont{Bohigas, Leboeuf,
  and Sánchez}}]{Bohigas1999}
\bibinfo{author}{\bibfnamefont{O.}~\bibnamefont{Bohigas}},
  \bibinfo{author}{\bibfnamefont{P.}~\bibnamefont{Leboeuf}}, \bibnamefont{and}
  \bibinfo{author}{\bibfnamefont{M.}~\bibnamefont{Sánchez}},
  \bibinfo{journal}{Physica D} \textbf{\bibinfo{volume}{131}},
  \bibinfo{pages}{186 } (\bibinfo{year}{1999}).

\bibitem[{\citenamefont{Casal et~al.}(2021)\citenamefont{Casal, Mu\~noz, and
  Molina}}]{Casal2021}
\bibinfo{author}{\bibfnamefont{I.}~\bibnamefont{Casal}},
  \bibinfo{author}{\bibfnamefont{L.}~\bibnamefont{Mu\~noz}}, \bibnamefont{and}
  \bibinfo{author}{\bibfnamefont{R.~A.} \bibnamefont{Molina}},
  \bibinfo{journal}{Eur. Phys. J. Plus} \textbf{\bibinfo{volume}{136}},
  \bibinfo{pages}{263} (\bibinfo{year}{2021}).

\bibitem[{\citenamefont{Che et~al.}(2021)\citenamefont{Che, Lu, Zhang, and
  Dietz}}]{Che2021}
\bibinfo{author}{\bibfnamefont{J.}~\bibnamefont{Che}},
  \bibinfo{author}{\bibfnamefont{J.}~\bibnamefont{Lu}},
  \bibinfo{author}{\bibfnamefont{X.}~\bibnamefont{Zhang}}, \bibnamefont{and}
  \bibinfo{author}{\bibfnamefont{B.}~\bibnamefont{Dietz}},
  \bibinfo{journal}{Phys.Rev.E} \textbf{\bibinfo{volume}{103}},
  \bibinfo{pages}{accepted} (\bibinfo{year}{2021}).

\end{thebibliography}
\end{document}